\documentclass[twocolumn,prd,longbibliography,
 amsmath,amssymb,superscriptaddress]{revtex4-2}
\usepackage{graphicx}
\usepackage{dcolumn}
\usepackage{bm}
\usepackage[colorlinks=true, citecolor=blue, linkcolor=blue, urlcolor=blue]{hyperref}
\usepackage{xcolor}
\usepackage{amsmath, amssymb, amsfonts}
\usepackage{mathtools}
\usepackage{subfigure}
\usepackage{mathrsfs}
\usepackage{sidecap}
\usepackage{verbatim}

\definecolor{purple1}{rgb}{128,0,128}

\usepackage{bigints}
\newcommand{\bea}{\begin{eqnarray}}
\newcommand{\ea}{\end{eqnarray}}

\definecolor{darkpastelgreen}{rgb}{0.01, 0.75, 0.24}

\def\x{\mathbf{x}}

\def\d{\mathrm{d}}

\begin{document}

\title{
Probing cosmic expansion through small anisotropies in a Bose-Einstein condensate analogue}
\author{Miguel \surname{Citeli de Freitas}}
\email{miguelciteli@hotmail.com}
\affiliation{International Center of Physics, Institute of Physics, University of Brasilia, 70297-400 Brasilia, Federal District, Brazil} 
\author{Tha\"is \surname{Nivaille}}
\email{thais.nivaille@gmail.com}
\affiliation{International Center of Physics, Institute of Physics, University of Brasilia, 70297-400 Brasilia, Federal District, Brazil} 
\affiliation{Namur Institute of Structured Matter, Department of Physics, University of Namur, 61 rue de Bruxelles, 5000 Namur, 
Belgium}

\author{Caio C. \surname{Holanda Ribeiro}}
\email{caiocesarribeiro@alumni.usp.br}
\affiliation{International Center of Physics, Institute of Physics, University of Brasilia, 70297-400 Brasilia, Federal District, Brazil} 

\date\today

\begin{abstract}

In this letter, it is shown that by measuring quantities like density variance in the present state of an analogue universe one can assess information about its history due to the mere existence of small anisotropies in earlier times, here modeled by non-uniform densities. This offers a valuable tool to probe and compare cosmic expansion models and, as an application, it is shown that quantum fluctuations can have a significant impact on the observable universe depending on its history.


\end{abstract}

\maketitle

\emph{Introduction.}---Bose-Einstein condensates (BEC) are versatile platforms to simulate curved space phenomena. The literature on analogue gravity in BECs is rich \cite{Unruh1981,Matt,Garay,Fischer2004,PhysRevA.78.021603,PhysRevLett.105.240401,Visser2011,PhysRevA.80.043601,Steinhauer2016,PhysRevA.79.043616,Florent2,PhysRevD.106.L061701}, and one of the most impressive milestones was the measurement of the analogue Hawking radiation \cite{Jeff2019,Jeff2021}. Further examples include measurements with analogue universes \cite{Jacobson2017,Steinhauer2022,Marius2024} and black holes with inner horizons \cite{Steinhauer2014}. Also, theoretical works involving BEC analogues reached goals that far superseded original motivations. Examples include quantum information aspects in black hole analogues \cite{PhysRevA.104.063302}, trans-Planckian effects in dipolar black hole analogues \cite{PhysRevD.107.L121502}, black hole lasers \cite{nova2016,Florent1,nova2023}, and analogue systems for the dynamical Casimir effect \cite{Westbrook2012}. Part of the scientific successes of analogue gravity in BECs is explained by the level of control in modeling such systems, as most analogues are based on diluted condensates with quantum fluctuations well described by the Bogoliubov expansion \cite{PhysRevA.80.043603,PhysRevA.85.013621,PhysRevD.97.025006}. In contrast, a full quantum description in other platforms can be unattainable. For instance, optical systems can have ill-defined quantum correlations even in simplified configurations \cite{PhysRevD.59.085011}.    


Another intricate problem of quantum field theory in curved space that can be modeled with BECs is the quantum backreaction. We recall that Hawking radiation is a semiclassical phenomenon \cite{HAWKING1974}, and to determine how the evaporation process affects the underlying black hole is a convoluted and rewarding problem. For instance, by assuming that semiclassical gravity remains valid throughout the evaporation process, radiation of primordial black holes may have an impact on the observable dark matter and other astrophysical quantities \cite{Cheek,AUFFINGER2023104040}. In this context, BECs can be used to study the quantum backreaction phenomena in analogue models. In fact, different methods exist in such a direction \cite{PhysRevLett.94.161302,PhysRevD.72.105005,Liberati2020}, with known exact solutions \cite{RibeiroPRA}, and applications to cosmological models \cite{PhysRevLett.130.241501}.     

Using as motivation the use of BEC to mimic expanding universes \cite{Weinfurtner2005,Jacobson2017,PhysRevLett.118.130404,Jain,zache2017,PhysRevA.100.043613,PhysRevA.104.023302,Mireia,PhysRevA.103.023322,PhysRevD.110.123523}, in this letter we show that small anisotropies in a 1D condensate can be used to test how distinct scale factors affect the current state of the universe, thereby inaugurating a route to assess the universe history in the laboratory. Specifically, we consider a non-relativistic gas of bosons tightly confined in an expanding ring-like configuration and we model anisotropies as small deviations from a uniform density. Therefore, anisotropies in this 1D gas are henceforth interpreted as density variations as function of the polar angle. By considering small quantum fluctuations on top of this expanding background ring, we show how each field mode can be linked to a fictitious particle in a Friedmann-Lemaître-Robertson-Walker (FLRW) spacetime \cite{bessa}, similarly to what was found in \cite{Jacobson2017}. Furthermore, we show that the small anisotropies give rise to an effective coupling between the fictitious particles, thereby producing an analogue model for interacting particles in a FLRW universe. Moreover, we shall see that once the initial distribution of anisotropies in the universe is known, by measuring certain quantities like the gas density variance, one can infer information about how the expansion occurred, which is a core problem in cosmology \cite{Zhengxiang,HUTERER201523,Moresco2022}. In fact, as an even stronger result, we show that two universes with the same  anisotropies at a given instant of time and that are evolving with similar scale factors can {\it become} rather different depending on their earlier histories, which we interpret as a sort of quantum memory closely related to quantum particle production in curved spaces \cite{Sandro,Birrell}.

In general, although condensates with fairly uniform density can be prepared \cite{Navon2021}, from an experimental perspective the assumption that a condensate realization has a uniform density is rather restrictive and we expect some defects (the anisotropies) to be present. Our analysis thus shows how to exploit such defects in studying the condensate dynamics and our findings depend only on their existence, and not on their magnitude. Conversely, the defects can always be excited before the condensate expansion takes place \cite{Jacobson2017}. 

\emph{The analogue universe.}---Our analogue model is based on a 3D boson gas subjected to a certain trapping potential that allows for an effective 1D gas description. The details of the dimensional reduction procedure are given in the supplementary material \cite{supp}. We adopt cylindrical coordinates and let the gas be in the $z=0$ plane, described by $\phi=\phi(t,\theta)$, $-\pi<\theta<\pi$, and we provide here only the important notation-simplifying conventions. The field $\phi$ is such that $\int_{-\pi}^\pi\d\theta|\phi|^2=N$, $N$ being the (conserved) number of particles in the gas.   In addition to $\hbar=1$, we let $g_0>0$ be a constant reference value for the interaction strength $g$, and $\mu_0=g_0N/\ell_0$ be the initial chemical potential in terms of ring initial length $\ell_0=2\pi R_0$. If $m$ is the mass of the gas particles, we henceforth express distances in units of $\xi_0=1/\sqrt{m\mu_0}$, $t$ in units of $1/\mu_0$, $g$ in units of $g_0$, and $\phi$ in units of $\sqrt{N}$ \cite{supp}. With these conventions, $\phi$ is subjected to the Gross-Pitaevskii equation \cite{supp}
\begin{equation}
    i\partial_t\phi=-\frac{\partial^2_\theta\phi}{2R^2}+\left(V+\frac{\ell_0}{R}g|\phi|^2\right)\phi,\label{mainfieldeq2}
\end{equation}
where $R=R(t)$ is the ring radius. 
%
%
Two points on the condensate separated by an angle $\theta$ are at a distance  of $d=\theta R$ from each other (along the gas). In particular, when $R$ changes over time, then so does this distance.

\begin{figure}
    \centering
    \includegraphics[width=1.\linewidth]{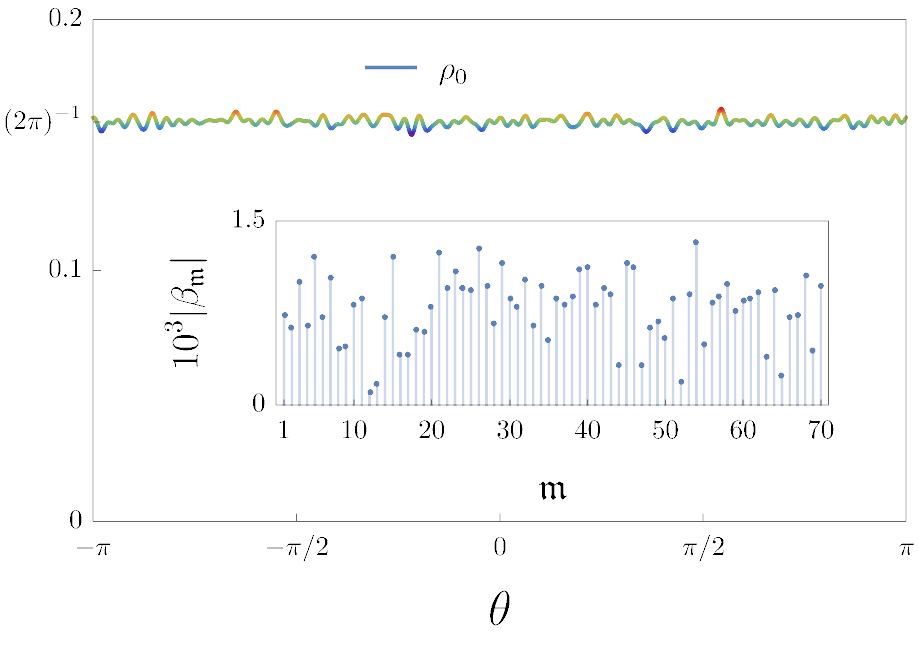}
    \caption{Condensate density (continuous curve) of a possible background condensate, which has the averaged value of $1/(2\pi)$. In this plot, the anisotropy coefficients $\beta_\mathfrak{m}$ were generated randomly and have absolute values shown in the inset.}
    \label{figdensity}
\end{figure}

Our analogue model assumes a background solution of Eq.~\eqref{mainfieldeq2} that has a time-independent density profile \cite{Mireia} and we take the condensate density to be ``almost'' uniform, i.e., the analogue universe is approximately homogeneous, with eventual deviations from uniformity being treated as corrections. Within the analogue gravity interpretation, this setup is such that anisotropies expand alongside the whole universe. Now, as the set $\{\exp(in\theta)\}$, with $n\in\mathbb{Z}$, is complete on the interval $[-\pi,\pi]$, the most general condensate density satisfying these assumptions must have the form
\begin{equation}
    \rho_0(\theta)=\frac{1}{2\pi}\left(1+\sum_{\mathfrak{m}=-\infty}^{\infty}\beta_\mathfrak{m}e^{i\mathfrak{m}\theta}\right),\label{backgrounddensity}
\end{equation}
where $|\beta_\mathfrak{m}|\ll1$. Also, we observe $\beta_{-\mathfrak{m}}=\beta^*_\mathfrak{m}$, and $\beta_0=0$.  
%
%
By setting the external potential as
\begin{align}
    V=1-&\frac{gR_0}{R}-\sum_{\mathfrak{m}}\beta_\mathfrak{m}\left(\frac{\mathfrak{m}^2}{4R^2}+\frac{gR_0}{R}\right)e^{i\mathfrak{m}\theta},
\end{align}
$\phi\equiv\phi_0=\sqrt{\rho_0}\exp(-i t)$ is a solution of Eq.~\eqref{mainfieldeq2}. We depict in Fig.~\ref{figdensity} an example of the possible anisotropies included in Eq.~\eqref{backgrounddensity}.

We are interested in studying small quantum effects on top of a condensate background that expands with time. This can be achieved, e.g., by means of the Bogoliubov expansion $\phi=\phi_0+\chi+\ldots$, where formally $|\chi|\ll|\phi_0|$. Thus, by ``linearizing'' Eq.~\eqref{mainfieldeq2}, we find that $\chi$ is ruled by the  
the Bogoliubov-de Gennes (BdG) equation \cite{RibeiroPRA}
\begin{align}
i\partial_t\chi=\left(-\frac{\partial_\theta^2}{2R^2}+V+2\frac{\ell_0}{R}g\rho_0\right)\chi+\frac{\ell_0}{R}g\phi_{\rm 0}^2\chi^{*}.\label{BdG}
\end{align}
We assume that the field $\chi$ encapsulates all information regarding the quantum features on top of the expanding background. We refer to \cite{castin} and references therein for a detailed account on the quantization of this field. For our purposes, it is enough to state that $\chi$ is to be promoted to an operator-valued distribution solution of Eq.~\eqref{BdG} satisfying canonical commutation relations \cite{supp}.
%
%
The quantum state of the condensate is here assumed to be any vacuum state fixed by suitable physical requirements, for which $\langle\chi\rangle=0$. 

Before we proceed to the presentation of our main result, it is instructive to show how an analogue gravity interpretation emerges for the system under study. In general, such an interpretation can be found by relating $\chi$ to density and phase perturbations \cite{Jacobson2017}. Equivalently, let us consider the variable $\psi(t,\theta)=\exp(i\mu t)\chi(t,\theta)$ and the Nambu spinor $\Psi=(\psi,\psi^*)^{T}$, where $T$ is the matrix transpose. The BdG equation can then be used to establish an analogy between the field $\Psi$ and particles in a FLRW spacetime.  In fact, $\Psi$ can always be written in terms of two scalar variables $f_n$ and $h_n$ as
\begin{equation}
    \Psi(t,\theta)=\sum_ne^{in\theta}\left[f_n(t)\left(\begin{array}{c}
         1  \\
         1 
    \end{array}\right)+h_n(t)\left(\begin{array}{c}
         1  \\
         -1 
    \end{array}\right)\right],\label{ansatzinterpretation}
\end{equation}
such that the BdG equation assumes the form
\begin{align}
    &i\dot{f}_n-\omega_nh_n=-\sum_{\mathfrak{m}}\frac{\beta_{\mathfrak{m}}}{2}\omega_\mathfrak{m}h_{n-\mathfrak{m}},\label{eqsystem1}\\
    &i\dot{h}_n-(\omega_n+\gamma)f_n=\sum_{\mathfrak{m}}\frac{\beta_\mathfrak{m}}{2}(2\gamma-\omega_\mathfrak{m})f_{n-\mathfrak{m}},\label{eqsystem2}
\end{align}
where
\begin{align}
    \omega_n&=\frac{n^2}{2R^2},\ \gamma=\frac{2R_0g}{R}.
\end{align}
By treating the system \eqref{eqsystem1}, \eqref{eqsystem2} perturbatively in $\beta_{\mathfrak{m}}$, we find, for $n\neq0$, that 
\begin{align}
    \ddot{f}_n+\frac{2\dot{{R}}}{R}&\dot{f}_n+\omega_n(\omega_n+\gamma)f_n=\nonumber\\
    =\sum_{\mathfrak{m}}\frac{\beta_\mathfrak{m}}{2}&[\omega_{\mathfrak{m}}(\omega_{n-\mathfrak{m}}+\gamma)-\omega_n(2\gamma-\omega_{\mathfrak{m}})]f_{n-\mathfrak{m}},
\end{align}
up to linear order in $\beta_\mathfrak{m}$. If we identify $\dot{f}_n$ with one of the components of the velocity of a fictitious particle, then when $\beta_\mathfrak{m}=0$ the equation above is precisely the geodesic equation for this particle in a FLRW universe with scale factor $a(t)=R(t)$ and subjected to the time-dependent potential $\omega_n(\omega_n+\gamma)$ \cite{bessa}. Thus, each mode in Eq.~\eqref{ansatzinterpretation} corresponds to a particle in a truly curved spacetime, and the anisotropies effectively couple such particles. Therefore, one can investigate the evolution of {\it interacting} particles in a FLRW universe by studying small perturbations (via $\chi$) on top of the background condensate.

\emph{Impact of the expansion on the BEC observables.}--- We now show that as the analogue universe expands, {\it any} initial anisotropy acts as a probe of the expansion.  Among the various quantities of interest that can be measured in condensates, we consider here one that is essential for the analysis of quantum backreaction \cite{PhysRevA.111.023306} and was recently used to probe the analogue Hawking radiation, namely, the local density variance, $G=\langle:(\rho-\langle\rho\rangle)^2:\rangle$, which is the diagonal part of the density-density correlation function \cite{Jeff2019}. We kept $G$ without the superscript in order to simplify the notation \footnote{Usually the density variance is denoted by $G^{(2)}$}. Here, $\rho=(\phi^*_0+\chi^\dagger)(\phi_0+\chi)$, and the double dots inside the VEV indicates normal ordering: all $\chi^\dagger$ are put to the left of all $\chi$ in monomials. This quantity is necessary, for instance, to determine the energy stored in the quantum fluctuations \cite{PhysRevA.111.023306}, which is a topic of fundamental importance in cosmology. 

A quantum field expansion for $\chi$, or, equivalently, for $\Psi$, can be found by standard canonical methods. We show in \cite{supp} that a complete set of positive norm field modes $\{\Psi_n\}_{n\in\mathbb{Z}}$, normalized to unit according to the BdG scalar product, can be found in the form
\begin{align}
    \Psi_n(t,\theta)=e^{in\theta}\Psi^{(0)}_n(t)+\sum_\mathfrak{m}\beta_{\mathfrak{m}}e^{i(n+\mathfrak{m})\theta}\Psi^{(1)}_{n,\mathfrak{m}}(t).\label{mainfieldmodes}
\end{align}
Therefore,
\begin{equation}
    \Psi=\sum_{n}(a_n\Psi_n+a_n^*\sigma_1\Psi_n^*),\label{mainexpansion}
\end{equation}
where $\sigma_i$, $i=1,2,3$ are the Pauli matrices. Quantization is thus finished by promoting each of the Fourier coefficients $a_n$ in Eq.~\eqref{mainexpansion} to an operator subjected to the commutation relation $[a_n,a^\dagger_{n'}]=\delta_{nn'}$, and we identify the state $|0\rangle$, defined by $a_n|0\rangle=0$, for all $n$, with the system vacuum state. If $\Psi$ is an arbitrary spinor, we denote its components by $\Psi_{i}$, $i=1,2$, and thus we obtain the quantum field expansion for $\chi$ by taking the first component of Eq.~\eqref{mainexpansion}:
\begin{equation}
    \chi(t,\theta)=e^{-i\mu t}\sum_{n\in\mathbb{Z}}\left[a_n\Psi_{n,1}(t,\theta)+a^\dagger_n\Psi^*_{n,2}(t,\theta)\right].\label{quantumfieldexpansion}
\end{equation}

Note that in the analogue condensate system there are two phenomena that break the system time translation symmetry. The first and obvious one is the fact that at $t=0$ external laboratory agents act on the gas in order to make the ring expand. The second source of time translation symmetry breakdown is a true quantum mechanical one: the condensate phase diffusion \cite{lew}. From the perspective of quantum field theory, this symmetry breakdown leads to a difficulty in assessing the condensate quantum (vacuum) state at a given time in order to compare it with theoretical predictions, in the same manner as occurs in cosmology \cite{Sandro}. To overcome this difficulty, we shall assume that the condensate history is known and it starts its evolution at $t=0$ from a stationary configuration, for which $R$ is constant and $g=0$. We cite \cite{PhysRevA.111.023306} for a recent account on the importance of this assumption when studying quantum fluctuations in condensates. 

We are now able to write down the major result of our analysis. In general, we find, by means of Eq.~\eqref{mainfieldmodes}, that
\begin{align}
    G(t,\theta)&=G^{(0)}(t)+\frac{1}{2\pi}\sum_{\mathfrak{m}}\beta_\mathfrak{m}G^{(1)}_{\mathfrak{m}}(t)e^{i\mathfrak{m}\theta},\label{ggeneral}
\end{align}
up to linear order in $\beta_{\mathfrak{m}}$, where  
\begin{align}
    &G^{(1)}_{\mathfrak{m}} =\sum_{n}\Big[\Psi^{(0)*}_{n,2}\Big(\Psi^{(0)}_{n,1}+\Psi^{(1)}_{n,\mathfrak{m},1}+\Psi^{(0)}_{n,2}+\Psi^{(1)}_{n,\mathfrak{m},2}\Big)\nonumber\\
    &+\Psi^{(0)}_{n,2}\Big(\Psi^{(0)*}_{n,1}+\Psi^{(1)*}_{n,-\mathfrak{m},1}+\Psi^{(0)*}_{n,2}+\Psi^{(1)*}_{n,-\mathfrak{m},2}\Big)\nonumber\\
    &+\Psi^{(1)*}_{n,-\mathfrak{m},2}\Big(\Psi^{(0)}_{n,1}+\Psi^{(0)}_{n,2}\Big)+\Psi^{(1)}_{n,\mathfrak{m},2}\Big(\Psi^{(0)*}_{n,1}+\Psi^{(0)*}_{n,2}\Big)\Big].
\end{align}
Therefore,  information about the analogue universe expansion can be obtained by studying the power spectrum of $G$, as the magnitudes of the initial anisotropies are modulated by $G^{(1)}_{\mathfrak{m}}$: $\beta_{\mathfrak{m}}G^{(1)}_{\mathfrak{m}}=\int\d\theta \exp(-i\mathfrak{m}\theta)G$.

It should be mentioned that finding analytical solutions for the field modes $\Psi_n$ might not be possible due to arbitrary dependence on time of both $\omega_n$ and $\gamma$ through $R(t),g(t)$. It turns out that there is a family of analogues for which the field modes assume a simple form for {\it arbitrary} $R(t)$. In fact, let us assume that after the particle interactions are turned on at $t=0$ and the condensate expansion starts to take place, $g=g(t)$ depends on the spatial location of the ring through $gR=R_0$. In this particular regime, $R^2\omega_n$ and $R^2\gamma$ remain time-independent, and, by working with the variable $\tau$ defined as
\begin{equation}
    \tau=\int_0^{t}\frac{\d t'}{R^2(t')},
\end{equation}
the BdG equation becomes stationary. Thus, the time marker $\tau$ plays the role of a ``conformal time'' for this system. The explicit forms of $\Psi^{(0)}_n$ and $\Psi^{(1)}_{n,\mathfrak{m}}$ [cf. Eq.~\eqref{mainfieldmodes}] are presented in \cite{supp}.

To illustrate how measurements of \eqref{ggeneral} can be used to probe cosmic expansion, let us consider two models: a bouncing universe, $R(t)\equiv R_{\rm par}(t)$, and an exponential expansion, $R(t)\equiv R_{\rm exp}(t)$, where
\begin{align}
    R_{\rm par}(t)&=(R_2-R_1)\left(\frac{t-t_1}{t_2-t_1}\right)^2+R_1,\\
    R_{\rm exp}(t)&=R_2\left(\frac{R_1}{R_2}\right)^{(t_2-t)/(t_2-t_1)},
\end{align}
and $0<t_1<t_2$, $0<R_1,R_2$ are parameters.
\begin{figure}
    \centering
    \includegraphics[width=1.\linewidth]{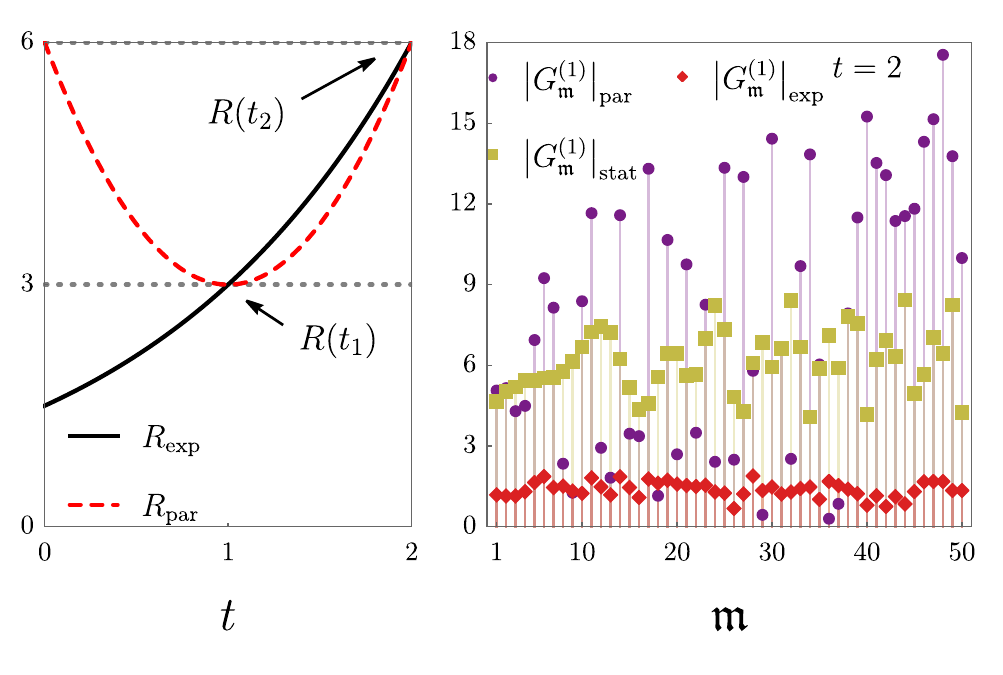}
    \includegraphics[width=0.96\linewidth]{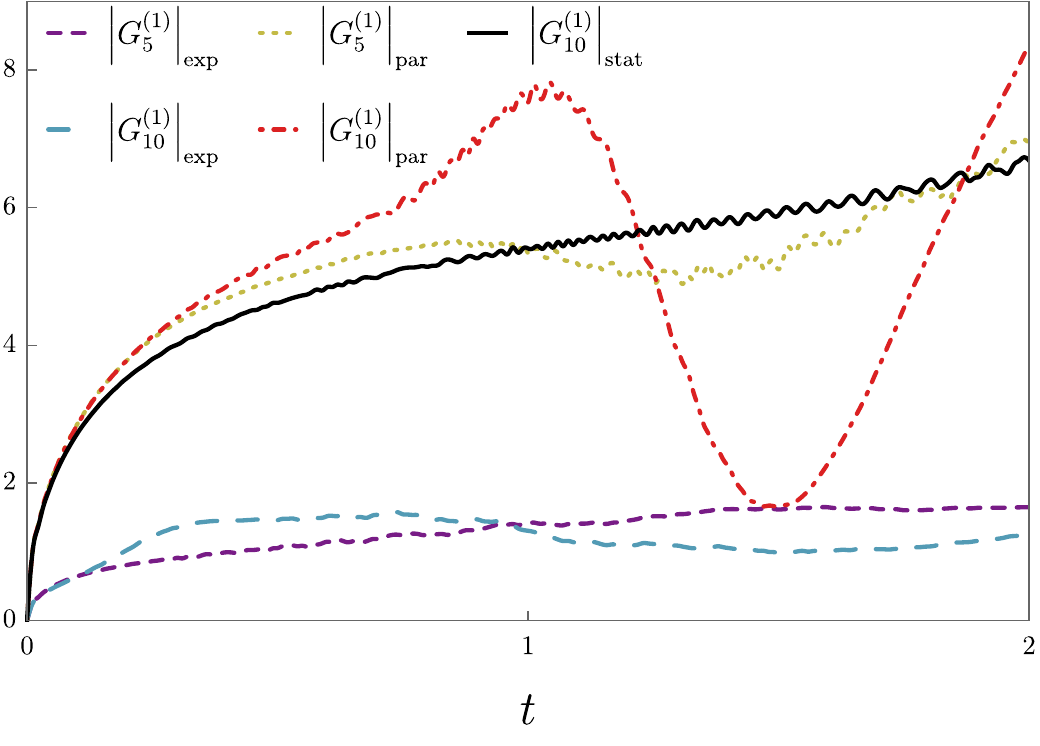}
    \caption{Upper left: Time-dependence of the condensate radius for two models, an exponential growth (black continuous curve), and a bouncing universe, with a parabolic (red dashed curve) scale factor. The two universes have the same size when $t=t_1,t_2$. Upper right: Magnitude of $G^{(1)}_{\mathfrak{m}}$ at $t=2$ for the three expansion models and several values of $\mathfrak{m}$. Note that each expansion leads to a unique profile. Bottom: Evolution of $|G^{(1)}_{\mathfrak{m}}|$ as function of time for the exponential ($|G^{(1)}_{\mathfrak{m}}|_{\rm exp}$), parabolic ($|G^{(1)}_{\mathfrak{m}}|_{\rm par}$), and static ($|G^{(1)}_{\mathfrak{m}}|_{\rm stat}$) universe models, for $\mathfrak{m}=5,10$. Note that although the three universes have the same scale factor at $t=2$, the magnitudes of the anisotropies are highly dependent on their histories. Also, even though the universes have the same size at $t=1$ and $t=2$, anisotropies of the same magnitude at $t=1$ can evolve very differently for $1<t<2$ depending on the early history of the universe.}
    \label{figmain}
\end{figure}
Figure \ref{figmain} upper left panel depicts the characteristic behavior of the two expansion models involving typical condensate sizes \cite{Steinhauer2014,Jeff2021}. Note that $R_{\rm exp}$ and $R_{\rm par}$ coincide at $t=t_1$ and $t=t_2$, and if $R_2>R_1$, the model $R_{\rm par}$ represents a bouncing universe with a scale factor that depends on time in a parabolic manner, with minimum size given by $R_1$ at $t=t_1$.

Given the two universe models and the general quantum field expansion, it is straightforward to determine how $G^{(1)}_{\mathfrak{m}}$ evolves, as shown in Fig.~\ref{figmain}. The upper right panel shows $|G_{\mathfrak{m}}^{(1)}|$ evaluated at $t=t_2$ for several values of $\mathfrak{m}$, for the parabolic, $|G_{\mathfrak{m}}^{(1)}|_{\rm par}$, and the exponential, $|G_{\mathfrak{m}}^{(1)}|_{\rm exp}$, universe models. We also consider $|G_{\mathfrak{m}}^{(1)}|=|G_{\mathfrak{m}}^{(1)}|_{\rm stat}$ for the static universe of constant size $R_f$. The figure shows that the $|G_{\mathfrak{m}}^{(1)}|$ combine to form a sort of fingerprint of the universe evolution. Note that the anisotropies depend on the particular experimental apparatus and conditions upon which the condensate is prepared, and by knowing the initial condensate density one can use the (usually unwanted) small anisotropies to study quantum fluctuations in the analogue universe. Also, this shows how a measurement of the density variance can be used to single out which expansion occurred.

Figure \ref{figmain} bottom panel  depicts $|G_{5}^{(1)}|_{\rm par},|G_{5}^{(1)}|_{\rm exp}, |G_{10}^{(1)}|_{\rm par},|G_{10}^{(1)}|_{\rm exp}$, and  $|G_{10}^{(1)}|_{\rm stat}$ as function of time, and it shows that the evolution of $G_{\mathfrak{m}}^{(1)}$ is strongly dependent on the underlying condensate history. Another striking feature revealed by the plots in Fig.~\ref{figmain} bottom panel, which is a sort of quantum memory effect stemming from earlier times, is how the amplitude of $G^{(1)}_{10}$ changes between $t=t_1$ and $t=t_2$ for the exponential and parabolic models. We recall that during this time period both universes modeled by $R_{\rm exp}$ and $R_{\rm par}$ are fairly similar, sharing the same size when $t=t_1,t_2$ and similar expansion rates. Yet, $G^{(1)}_{10}$ decreases by a factor of approximately $8$ for the parabolic universe during this period, whereas for the exponentially expanding condensate it remains approximately constant. We attribute this memory effect to the fact that although both universes have similar expansions during $t_1<t<t_2$, their quantum vacua are distinct at $t=t_1$ due to quantum particle production at earlier times \cite{Birrell,Sandro}.

\emph{Final remarks.}--- In this work, we showed how small deviations from a uniform condensate density lead to a remarkable memory effect in expanding universe analogues that can be effectively used as probes of the expansion dynamics. The key ideas in our model are traced back into the main hypotheses: anisotropies, to some degree, are expected to be present in experimental realizations of BECs and in the case of expanding universe analogues they must leave imprints on the observable quantities. As a consequence of these ideas, we showed how the existence of small anisotropies, that are usually neglected in theoretical investigations, gives rise to an interesting tool to test different expansion models. 

Before we finish this analysis, a few important remarks are in order. Although we restricted our argument to the density variance, we note that all observables that are quadratic in the field variables will have the same general form of Eq.~\eqref{ggeneral}. This includes, for instance, other measurable quantities like quantum depletion \cite{lopes}, energy density, and their fluxes, which appear in the study of quantum backreaction \cite{PhysRevA.111.023306}. Our findings suggest that, depending on their evolution, quantum fluctuations in the early universe can have rather significant effects at present times via the enhancement or suppression of initial anisotropies, thus showing that a quantum backreaction analysis can unveil interesting new phenomena in both the analogue universe and in curved spaces.

Finally, we presented an explicit result for a universe analogue for which $gR=R_0$. This assumption has the advantage of avoiding numerical methods, at the expense of assuming a coupling constant that varies with the position of the ring. Nevertheless, our findings do not depend on this particular assumption, to the extent that Eq.~\eqref{ggeneral} maintains its form for {\it all} condensate configurations and quantum memory effects induced by the ring expansion will be present regardless of the particular form of $g$.

\emph{Acknowledgements.}---
M.C.F. acknowledges a support from the Brazilian funding agency CAPES (Grant N.
88887.007464/2024-00). C.C.H.R. would like to thank the Funda\c{c}\~ao de Apoio \`a Pesquisa do Distrito
Federal (Grant N. 00193-00002051/2023-14) for supporting this work. 

\bibliography{qgav3.bib}

\begin{thebibliography}{58}%
\makeatletter
\providecommand \@ifxundefined [1]{%
 \@ifx{#1\undefined}
}%
\providecommand \@ifnum [1]{%
 \ifnum #1\expandafter \@firstoftwo
 \else \expandafter \@secondoftwo
 \fi
}%
\providecommand \@ifx [1]{%
 \ifx #1\expandafter \@firstoftwo
 \else \expandafter \@secondoftwo
 \fi
}%
\providecommand \natexlab [1]{#1}%
\providecommand \enquote  [1]{``#1''}%
\providecommand \bibnamefont  [1]{#1}%
\providecommand \bibfnamefont [1]{#1}%
\providecommand \citenamefont [1]{#1}%
\providecommand \href@noop [0]{\@secondoftwo}%
\providecommand \href [0]{\begingroup \@sanitize@url \@href}%
\providecommand \@href[1]{\@@startlink{#1}\@@href}%
\providecommand \@@href[1]{\endgroup#1\@@endlink}%
\providecommand \@sanitize@url [0]{\catcode `\\12\catcode `\$12\catcode
  `\&12\catcode `\#12\catcode `\^12\catcode `\_12\catcode `\%12\relax}%
\providecommand \@@startlink[1]{}%
\providecommand \@@endlink[0]{}%
\providecommand \url  [0]{\begingroup\@sanitize@url \@url }%
\providecommand \@url [1]{\endgroup\@href {#1}{\urlprefix }}%
\providecommand \urlprefix  [0]{URL }%
\providecommand \Eprint [0]{\href }%
\providecommand \doibase [0]{https://doi.org/}%
\providecommand \selectlanguage [0]{\@gobble}%
\providecommand \bibinfo  [0]{\@secondoftwo}%
\providecommand \bibfield  [0]{\@secondoftwo}%
\providecommand \translation [1]{[#1]}%
\providecommand \BibitemOpen [0]{}%
\providecommand \bibitemStop [0]{}%
\providecommand \bibitemNoStop [0]{.\EOS\space}%
\providecommand \EOS [0]{\spacefactor3000\relax}%
\providecommand \BibitemShut  [1]{\csname bibitem#1\endcsname}%
\let\auto@bib@innerbib\@empty
\bibitem [{\citenamefont {Unruh}(1981)}]{Unruh1981}%
  \BibitemOpen
  \bibfield  {author} {\bibinfo {author} {\bibfnamefont {W.~G.}\ \bibnamefont
  {Unruh}},\ }\bibfield  {title} {\bibinfo {title} {{Experimental Black-Hole
  Evaporation?}},\ }\href {https://doi.org/10.1103/PhysRevLett.46.1351}
  {\bibfield  {journal} {\bibinfo  {journal} {Phys. Rev. Lett.}\ }\textbf
  {\bibinfo {volume} {46}},\ \bibinfo {pages} {1351} (\bibinfo {year}
  {1981})}\BibitemShut {NoStop}%
\bibitem [{\citenamefont {Visser}(1998)}]{Matt}%
  \BibitemOpen
  \bibfield  {author} {\bibinfo {author} {\bibfnamefont {M.}~\bibnamefont
  {Visser}},\ }\bibfield  {title} {\bibinfo {title} {{Acoustic black holes:
  horizons, ergospheres and Hawking radiation}},\ }\href
  {https://doi.org/10.1088/0264-9381/15/6/024} {\bibfield  {journal} {\bibinfo
  {journal} {Classical and Quantum Gravity}\ }\textbf {\bibinfo {volume}
  {15}},\ \bibinfo {pages} {1767} (\bibinfo {year} {1998})}\BibitemShut
  {NoStop}%
\bibitem [{\citenamefont {Garay}\ \emph {et~al.}(2000)\citenamefont {Garay},
  \citenamefont {Anglin}, \citenamefont {Cirac},\ and\ \citenamefont
  {Zoller}}]{Garay}%
  \BibitemOpen
  \bibfield  {author} {\bibinfo {author} {\bibfnamefont {L.~J.}\ \bibnamefont
  {Garay}}, \bibinfo {author} {\bibfnamefont {J.~R.}\ \bibnamefont {Anglin}},
  \bibinfo {author} {\bibfnamefont {J.~I.}\ \bibnamefont {Cirac}},\ and\
  \bibinfo {author} {\bibfnamefont {P.}~\bibnamefont {Zoller}},\ }\bibfield
  {title} {\bibinfo {title} {{Sonic Analog of Gravitational Black Holes in
  Bose-Einstein Condensates}},\ }\href
  {https://doi.org/10.1103/PhysRevLett.85.4643} {\bibfield  {journal} {\bibinfo
   {journal} {Phys. Rev. Lett.}\ }\textbf {\bibinfo {volume} {85}},\ \bibinfo
  {pages} {4643} (\bibinfo {year} {2000})}\BibitemShut {NoStop}%
\bibitem [{\citenamefont {Fischer}\ and\ \citenamefont
  {Sch\"utzhold}(2004)}]{Fischer2004}%
  \BibitemOpen
  \bibfield  {author} {\bibinfo {author} {\bibfnamefont {U.~R.}\ \bibnamefont
  {Fischer}}\ and\ \bibinfo {author} {\bibfnamefont {R.}~\bibnamefont
  {Sch\"utzhold}},\ }\bibfield  {title} {\bibinfo {title} {{Quantum simulation
  of cosmic inflation in two-component Bose-Einstein condensates}},\ }\href
  {https://doi.org/10.1103/PhysRevA.70.063615} {\bibfield  {journal} {\bibinfo
  {journal} {Phys. Rev. A}\ }\textbf {\bibinfo {volume} {70}},\ \bibinfo
  {pages} {063615} (\bibinfo {year} {2004})}\BibitemShut {NoStop}%
\bibitem [{\citenamefont {Balbinot}\ \emph {et~al.}(2008)\citenamefont
  {Balbinot}, \citenamefont {Fabbri}, \citenamefont {Fagnocchi}, \citenamefont
  {Recati},\ and\ \citenamefont {Carusotto}}]{PhysRevA.78.021603}%
  \BibitemOpen
  \bibfield  {author} {\bibinfo {author} {\bibfnamefont {R.}~\bibnamefont
  {Balbinot}}, \bibinfo {author} {\bibfnamefont {A.}~\bibnamefont {Fabbri}},
  \bibinfo {author} {\bibfnamefont {S.}~\bibnamefont {Fagnocchi}}, \bibinfo
  {author} {\bibfnamefont {A.}~\bibnamefont {Recati}},\ and\ \bibinfo {author}
  {\bibfnamefont {I.}~\bibnamefont {Carusotto}},\ }\bibfield  {title} {\bibinfo
  {title} {{Nonlocal density correlations as a signature of Hawking radiation
  from acoustic black holes}},\ }\href
  {https://doi.org/10.1103/PhysRevA.78.021603} {\bibfield  {journal} {\bibinfo
  {journal} {Phys. Rev. A}\ }\textbf {\bibinfo {volume} {78}},\ \bibinfo
  {pages} {021603} (\bibinfo {year} {2008})}\BibitemShut {NoStop}%
\bibitem [{\citenamefont {Lahav}\ \emph {et~al.}(2010)\citenamefont {Lahav},
  \citenamefont {Itah}, \citenamefont {Blumkin}, \citenamefont {Gordon},
  \citenamefont {Rinott}, \citenamefont {Zayats},\ and\ \citenamefont
  {Steinhauer}}]{PhysRevLett.105.240401}%
  \BibitemOpen
  \bibfield  {author} {\bibinfo {author} {\bibfnamefont {O.}~\bibnamefont
  {Lahav}}, \bibinfo {author} {\bibfnamefont {A.}~\bibnamefont {Itah}},
  \bibinfo {author} {\bibfnamefont {A.}~\bibnamefont {Blumkin}}, \bibinfo
  {author} {\bibfnamefont {C.}~\bibnamefont {Gordon}}, \bibinfo {author}
  {\bibfnamefont {S.}~\bibnamefont {Rinott}}, \bibinfo {author} {\bibfnamefont
  {A.}~\bibnamefont {Zayats}},\ and\ \bibinfo {author} {\bibfnamefont
  {J.}~\bibnamefont {Steinhauer}},\ }\bibfield  {title} {\bibinfo {title}
  {{Realization of a Sonic Black Hole Analog in a Bose-Einstein Condensate}},\
  }\href {https://doi.org/10.1103/PhysRevLett.105.240401} {\bibfield  {journal}
  {\bibinfo  {journal} {Phys. Rev. Lett.}\ }\textbf {\bibinfo {volume} {105}},\
  \bibinfo {pages} {240401} (\bibinfo {year} {2010})}\BibitemShut {NoStop}%
\bibitem [{\citenamefont {Barcel{\'o}}\ \emph {et~al.}(2011)\citenamefont
  {Barcel{\'o}}, \citenamefont {Liberati},\ and\ \citenamefont
  {Visser}}]{Visser2011}%
  \BibitemOpen
  \bibfield  {author} {\bibinfo {author} {\bibfnamefont {C.}~\bibnamefont
  {Barcel{\'o}}}, \bibinfo {author} {\bibfnamefont {S.}~\bibnamefont
  {Liberati}},\ and\ \bibinfo {author} {\bibfnamefont {M.}~\bibnamefont
  {Visser}},\ }\bibfield  {title} {\bibinfo {title} {{Analogue Gravity}},\
  }\href {https://doi.org/10.12942/lrr-2011-3} {\bibfield  {journal} {\bibinfo
  {journal} {Living Reviews in Relativity}\ }\textbf {\bibinfo {volume} {14}},\
  \bibinfo {pages} {3} (\bibinfo {year} {2011})}\BibitemShut {NoStop}%
\bibitem [{\citenamefont {Macher}\ and\ \citenamefont
  {Parentani}(2009)}]{PhysRevA.80.043601}%
  \BibitemOpen
  \bibfield  {author} {\bibinfo {author} {\bibfnamefont {J.}~\bibnamefont
  {Macher}}\ and\ \bibinfo {author} {\bibfnamefont {R.}~\bibnamefont
  {Parentani}},\ }\bibfield  {title} {\bibinfo {title} {{Black-hole radiation
  in Bose-Einstein condensates}},\ }\href
  {https://doi.org/10.1103/PhysRevA.80.043601} {\bibfield  {journal} {\bibinfo
  {journal} {Phys. Rev. A}\ }\textbf {\bibinfo {volume} {80}},\ \bibinfo
  {pages} {043601} (\bibinfo {year} {2009})}\BibitemShut {NoStop}%
\bibitem [{\citenamefont {Steinhauer}(2016)}]{Steinhauer2016}%
  \BibitemOpen
  \bibfield  {author} {\bibinfo {author} {\bibfnamefont {J.}~\bibnamefont
  {Steinhauer}},\ }\bibfield  {title} {\bibinfo {title} {{Observation of
  quantum Hawking radiation and its entanglement in an analogue black hole}},\
  }\href {https://doi.org/10.1038/nphys3863} {\bibfield  {journal} {\bibinfo
  {journal} {Nature Physics}\ }\textbf {\bibinfo {volume} {12}},\ \bibinfo
  {pages} {959} (\bibinfo {year} {2016})}\BibitemShut {NoStop}%
\bibitem [{\citenamefont {Kurita}\ \emph {et~al.}(2009)\citenamefont {Kurita},
  \citenamefont {Kobayashi}, \citenamefont {Morinari}, \citenamefont
  {Tsubota},\ and\ \citenamefont {Ishihara}}]{PhysRevA.79.043616}%
  \BibitemOpen
  \bibfield  {author} {\bibinfo {author} {\bibfnamefont {Y.}~\bibnamefont
  {Kurita}}, \bibinfo {author} {\bibfnamefont {M.}~\bibnamefont {Kobayashi}},
  \bibinfo {author} {\bibfnamefont {T.}~\bibnamefont {Morinari}}, \bibinfo
  {author} {\bibfnamefont {M.}~\bibnamefont {Tsubota}},\ and\ \bibinfo {author}
  {\bibfnamefont {H.}~\bibnamefont {Ishihara}},\ }\bibfield  {title} {\bibinfo
  {title} {{Spacetime analog of Bose-Einstein condensates: Bogoliubov--de
  Gennes formulation}},\ }\href {https://doi.org/10.1103/PhysRevA.79.043616}
  {\bibfield  {journal} {\bibinfo  {journal} {Phys. Rev. A}\ }\textbf {\bibinfo
  {volume} {79}},\ \bibinfo {pages} {043616} (\bibinfo {year}
  {2009})}\BibitemShut {NoStop}%
\bibitem [{\citenamefont {Michel}\ and\ \citenamefont
  {Parentani}(2015)}]{Florent2}%
  \BibitemOpen
  \bibfield  {author} {\bibinfo {author} {\bibfnamefont {F.}~\bibnamefont
  {Michel}}\ and\ \bibinfo {author} {\bibfnamefont {R.}~\bibnamefont
  {Parentani}},\ }\bibfield  {title} {\bibinfo {title} {{Nonlinear effects in
  time-dependent transonic flows: An analysis of analog black hole
  stability}},\ }\href {https://doi.org/10.1103/PhysRevA.91.053603} {\bibfield
  {journal} {\bibinfo  {journal} {Phys. Rev. A}\ }\textbf {\bibinfo {volume}
  {91}},\ \bibinfo {pages} {053603} (\bibinfo {year} {2015})}\BibitemShut
  {NoStop}%
\bibitem [{\citenamefont {Tian}\ \emph {et~al.}(2022)\citenamefont {Tian},
  \citenamefont {Wu}, \citenamefont {Zhang}, \citenamefont {Jing},\ and\
  \citenamefont {Du}}]{PhysRevD.106.L061701}%
  \BibitemOpen
  \bibfield  {author} {\bibinfo {author} {\bibfnamefont {Z.}~\bibnamefont
  {Tian}}, \bibinfo {author} {\bibfnamefont {L.}~\bibnamefont {Wu}}, \bibinfo
  {author} {\bibfnamefont {L.}~\bibnamefont {Zhang}}, \bibinfo {author}
  {\bibfnamefont {J.}~\bibnamefont {Jing}},\ and\ \bibinfo {author}
  {\bibfnamefont {J.}~\bibnamefont {Du}},\ }\bibfield  {title} {\bibinfo
  {title} {{Probing Lorentz-invariance-violation-induced nonthermal Unruh
  effect in quasi-two-dimensional dipolar condensates}},\ }\href
  {https://doi.org/10.1103/PhysRevD.106.L061701} {\bibfield  {journal}
  {\bibinfo  {journal} {Phys. Rev. D}\ }\textbf {\bibinfo {volume} {106}},\
  \bibinfo {pages} {L061701} (\bibinfo {year} {2022})}\BibitemShut {NoStop}%
\bibitem [{\citenamefont {de~Nova}\ \emph {et~al.}(2019)\citenamefont
  {de~Nova}, \citenamefont {Golubkov}, \citenamefont {Kolobov},\ and\
  \citenamefont {Steinhauer}}]{Jeff2019}%
  \BibitemOpen
  \bibfield  {author} {\bibinfo {author} {\bibfnamefont {J.~R.~M.}\
  \bibnamefont {de~Nova}}, \bibinfo {author} {\bibfnamefont {K.}~\bibnamefont
  {Golubkov}}, \bibinfo {author} {\bibfnamefont {V.~I.}\ \bibnamefont
  {Kolobov}},\ and\ \bibinfo {author} {\bibfnamefont {J.}~\bibnamefont
  {Steinhauer}},\ }\bibfield  {title} {\bibinfo {title} {{Observation of
  thermal Hawking radiation and its temperature in an analogue black hole}},\
  }\href {https://doi.org/10.1038/s41586-019-1241-0} {\bibfield  {journal}
  {\bibinfo  {journal} {Nature}\ }\textbf {\bibinfo {volume} {569}},\ \bibinfo
  {pages} {688} (\bibinfo {year} {2019})}\BibitemShut {NoStop}%
\bibitem [{\citenamefont {Kolobov}\ \emph {et~al.}(2021)\citenamefont
  {Kolobov}, \citenamefont {Golubkov}, \citenamefont {de~Nova},\ and\
  \citenamefont {Steinhauer}}]{Jeff2021}%
  \BibitemOpen
  \bibfield  {author} {\bibinfo {author} {\bibfnamefont {V.~I.}\ \bibnamefont
  {Kolobov}}, \bibinfo {author} {\bibfnamefont {K.}~\bibnamefont {Golubkov}},
  \bibinfo {author} {\bibfnamefont {J.~R.~M.}\ \bibnamefont {de~Nova}},\ and\
  \bibinfo {author} {\bibfnamefont {J.}~\bibnamefont {Steinhauer}},\ }\bibfield
   {title} {\bibinfo {title} {{Observation of stationary spontaneous Hawking
  radiation and the time evolution of an analogue black hole}},\ }\href
  {https://doi.org/10.1038/s41567-020-01076-0} {\bibfield  {journal} {\bibinfo
  {journal} {Nature Physics}\ }\textbf {\bibinfo {volume} {17}},\ \bibinfo
  {pages} {362} (\bibinfo {year} {2021})}\BibitemShut {NoStop}%
\bibitem [{\citenamefont {Eckel}\ \emph {et~al.}(2018)\citenamefont {Eckel},
  \citenamefont {Kumar}, \citenamefont {Jacobson}, \citenamefont {Spielman},\
  and\ \citenamefont {Campbell}}]{Jacobson2017}%
  \BibitemOpen
  \bibfield  {author} {\bibinfo {author} {\bibfnamefont {S.}~\bibnamefont
  {Eckel}}, \bibinfo {author} {\bibfnamefont {A.}~\bibnamefont {Kumar}},
  \bibinfo {author} {\bibfnamefont {T.}~\bibnamefont {Jacobson}}, \bibinfo
  {author} {\bibfnamefont {I.~B.}\ \bibnamefont {Spielman}},\ and\ \bibinfo
  {author} {\bibfnamefont {G.~K.}\ \bibnamefont {Campbell}},\ }\bibfield
  {title} {\bibinfo {title} {{A Rapidly Expanding Bose-Einstein Condensate: An
  Expanding Universe in the Lab}},\ }\href
  {https://doi.org/10.1103/PhysRevX.8.021021} {\bibfield  {journal} {\bibinfo
  {journal} {Phys. Rev. X}\ }\textbf {\bibinfo {volume} {8}},\ \bibinfo {pages}
  {021021} (\bibinfo {year} {2018})}\BibitemShut {NoStop}%
\bibitem [{\citenamefont {Steinhauer}\ \emph {et~al.}(2022)\citenamefont
  {Steinhauer}, \citenamefont {Abuzarli}, \citenamefont {Aladjidi},
  \citenamefont {Bienaim{\'e}}, \citenamefont {Piekarski}, \citenamefont {Liu},
  \citenamefont {Giacobino}, \citenamefont {Bramati},\ and\ \citenamefont
  {Glorieux}}]{Steinhauer2022}%
  \BibitemOpen
  \bibfield  {author} {\bibinfo {author} {\bibfnamefont {J.}~\bibnamefont
  {Steinhauer}}, \bibinfo {author} {\bibfnamefont {M.}~\bibnamefont
  {Abuzarli}}, \bibinfo {author} {\bibfnamefont {T.}~\bibnamefont {Aladjidi}},
  \bibinfo {author} {\bibfnamefont {T.}~\bibnamefont {Bienaim{\'e}}}, \bibinfo
  {author} {\bibfnamefont {C.}~\bibnamefont {Piekarski}}, \bibinfo {author}
  {\bibfnamefont {W.}~\bibnamefont {Liu}}, \bibinfo {author} {\bibfnamefont
  {E.}~\bibnamefont {Giacobino}}, \bibinfo {author} {\bibfnamefont
  {A.}~\bibnamefont {Bramati}},\ and\ \bibinfo {author} {\bibfnamefont
  {Q.}~\bibnamefont {Glorieux}},\ }\bibfield  {title} {\bibinfo {title}
  {{Analogue cosmological particle creation in an ultracold quantum fluid of
  light}},\ }\href {https://doi.org/10.1038/s41467-022-30603-1} {\bibfield
  {journal} {\bibinfo  {journal} {Nature Communications}\ }\textbf {\bibinfo
  {volume} {13}},\ \bibinfo {pages} {2890} (\bibinfo {year}
  {2022})}\BibitemShut {NoStop}%
\bibitem [{\citenamefont {Sparn}\ \emph {et~al.}(2024)\citenamefont {Sparn},
  \citenamefont {Kath}, \citenamefont {Liebster}, \citenamefont {Duchene},
  \citenamefont {Schmidt}, \citenamefont {Tolosa-Sime\'on}, \citenamefont
  {Parra-L\'opez}, \citenamefont {Floerchinger}, \citenamefont {Strobel},\ and\
  \citenamefont {Oberthaler}}]{Marius2024}%
  \BibitemOpen
  \bibfield  {author} {\bibinfo {author} {\bibfnamefont {M.}~\bibnamefont
  {Sparn}}, \bibinfo {author} {\bibfnamefont {E.}~\bibnamefont {Kath}},
  \bibinfo {author} {\bibfnamefont {N.}~\bibnamefont {Liebster}}, \bibinfo
  {author} {\bibfnamefont {J.}~\bibnamefont {Duchene}}, \bibinfo {author}
  {\bibfnamefont {C.~F.}\ \bibnamefont {Schmidt}}, \bibinfo {author}
  {\bibfnamefont {M.}~\bibnamefont {Tolosa-Sime\'on}}, \bibinfo {author}
  {\bibfnamefont {A.}~\bibnamefont {Parra-L\'opez}}, \bibinfo {author}
  {\bibfnamefont {S.}~\bibnamefont {Floerchinger}}, \bibinfo {author}
  {\bibfnamefont {H.}~\bibnamefont {Strobel}},\ and\ \bibinfo {author}
  {\bibfnamefont {M.~K.}\ \bibnamefont {Oberthaler}},\ }\bibfield  {title}
  {\bibinfo {title} {{Experimental Particle Production in Time-Dependent
  Spacetimes: A One-Dimensional Scattering Problem}},\ }\href
  {https://doi.org/10.1103/PhysRevLett.133.260201} {\bibfield  {journal}
  {\bibinfo  {journal} {Phys. Rev. Lett.}\ }\textbf {\bibinfo {volume} {133}},\
  \bibinfo {pages} {260201} (\bibinfo {year} {2024})}\BibitemShut {NoStop}%
\bibitem [{\citenamefont {Steinhauer}(2014)}]{Steinhauer2014}%
  \BibitemOpen
  \bibfield  {author} {\bibinfo {author} {\bibfnamefont {J.}~\bibnamefont
  {Steinhauer}},\ }\bibfield  {title} {\bibinfo {title} {{Observation of
  self-amplifying Hawking radiation in an analogue black-hole laser}},\ }\href
  {https://doi.org/10.1038/nphys3104} {\bibfield  {journal} {\bibinfo
  {journal} {Nature Physics}\ }\textbf {\bibinfo {volume} {10}},\ \bibinfo
  {pages} {864} (\bibinfo {year} {2014})}\BibitemShut {NoStop}%
\bibitem [{\citenamefont {Isoard}\ \emph {et~al.}(2021)\citenamefont {Isoard},
  \citenamefont {Milazzo}, \citenamefont {Pavloff},\ and\ \citenamefont
  {Giraud}}]{PhysRevA.104.063302}%
  \BibitemOpen
  \bibfield  {author} {\bibinfo {author} {\bibfnamefont {M.}~\bibnamefont
  {Isoard}}, \bibinfo {author} {\bibfnamefont {N.}~\bibnamefont {Milazzo}},
  \bibinfo {author} {\bibfnamefont {N.}~\bibnamefont {Pavloff}},\ and\ \bibinfo
  {author} {\bibfnamefont {O.}~\bibnamefont {Giraud}},\ }\bibfield  {title}
  {\bibinfo {title} {{Bipartite and tripartite entanglement in a Bose-Einstein
  acoustic black hole}},\ }\href {https://doi.org/10.1103/PhysRevA.104.063302}
  {\bibfield  {journal} {\bibinfo  {journal} {Phys. Rev. A}\ }\textbf {\bibinfo
  {volume} {104}},\ \bibinfo {pages} {063302} (\bibinfo {year}
  {2021})}\BibitemShut {NoStop}%
\bibitem [{\citenamefont {Holanda~Ribeiro}\ and\ \citenamefont
  {Fischer}(2023)}]{PhysRevD.107.L121502}%
  \BibitemOpen
  \bibfield  {author} {\bibinfo {author} {\bibfnamefont {C.~C.}\ \bibnamefont
  {Holanda~Ribeiro}}\ and\ \bibinfo {author} {\bibfnamefont {U.~R.}\
  \bibnamefont {Fischer}},\ }\bibfield  {title} {\bibinfo {title} {{Impact of
  trans-Planckian excitations on black-hole radiation in dipolar
  condensates}},\ }\href {https://doi.org/10.1103/PhysRevD.107.L121502}
  {\bibfield  {journal} {\bibinfo  {journal} {Phys. Rev. D}\ }\textbf {\bibinfo
  {volume} {107}},\ \bibinfo {pages} {L121502} (\bibinfo {year}
  {2023})}\BibitemShut {NoStop}%
\bibitem [{\citenamefont {de~Nova}\ \emph {et~al.}(2016)\citenamefont
  {de~Nova}, \citenamefont {Finazzi},\ and\ \citenamefont
  {Carusotto}}]{nova2016}%
  \BibitemOpen
  \bibfield  {author} {\bibinfo {author} {\bibfnamefont {J.~R.~M.}\
  \bibnamefont {de~Nova}}, \bibinfo {author} {\bibfnamefont {S.}~\bibnamefont
  {Finazzi}},\ and\ \bibinfo {author} {\bibfnamefont {I.}~\bibnamefont
  {Carusotto}},\ }\bibfield  {title} {\bibinfo {title} {{Time-dependent study
  of a black-hole laser in a flowing atomic condensate}},\ }\href
  {https://doi.org/10.1103/PhysRevA.94.043616} {\bibfield  {journal} {\bibinfo
  {journal} {Phys. Rev. A}\ }\textbf {\bibinfo {volume} {94}},\ \bibinfo
  {pages} {043616} (\bibinfo {year} {2016})}\BibitemShut {NoStop}%
\bibitem [{\citenamefont {Michel}\ and\ \citenamefont
  {Parentani}(2013)}]{Florent1}%
  \BibitemOpen
  \bibfield  {author} {\bibinfo {author} {\bibfnamefont {F.}~\bibnamefont
  {Michel}}\ and\ \bibinfo {author} {\bibfnamefont {R.}~\bibnamefont
  {Parentani}},\ }\bibfield  {title} {\bibinfo {title} {{Saturation of black
  hole lasers in Bose-Einstein condensates}},\ }\href
  {https://doi.org/10.1103/PhysRevD.88.125012} {\bibfield  {journal} {\bibinfo
  {journal} {Phys. Rev. D}\ }\textbf {\bibinfo {volume} {88}},\ \bibinfo
  {pages} {125012} (\bibinfo {year} {2013})}\BibitemShut {NoStop}%
\bibitem [{\citenamefont {de~Nova}\ and\ \citenamefont
  {Sols}(2023)}]{nova2023}%
  \BibitemOpen
  \bibfield  {author} {\bibinfo {author} {\bibfnamefont {J.~R.~M.}\
  \bibnamefont {de~Nova}}\ and\ \bibinfo {author} {\bibfnamefont
  {F.}~\bibnamefont {Sols}},\ }\bibfield  {title} {\bibinfo {title}
  {{Black-hole laser to Bogoliubov-Cherenkov-Landau crossover: From nonlinear
  to linear quantum amplification}},\ }\href
  {https://doi.org/10.1103/PhysRevResearch.5.043282} {\bibfield  {journal}
  {\bibinfo  {journal} {Phys. Rev. Res.}\ }\textbf {\bibinfo {volume} {5}},\
  \bibinfo {pages} {043282} (\bibinfo {year} {2023})}\BibitemShut {NoStop}%
\bibitem [{\citenamefont {Jaskula}\ \emph {et~al.}(2012)\citenamefont
  {Jaskula}, \citenamefont {Partridge}, \citenamefont {Bonneau}, \citenamefont
  {Lopes}, \citenamefont {Ruaudel}, \citenamefont {Boiron},\ and\ \citenamefont
  {Westbrook}}]{Westbrook2012}%
  \BibitemOpen
  \bibfield  {author} {\bibinfo {author} {\bibfnamefont {J.-C.}\ \bibnamefont
  {Jaskula}}, \bibinfo {author} {\bibfnamefont {G.~B.}\ \bibnamefont
  {Partridge}}, \bibinfo {author} {\bibfnamefont {M.}~\bibnamefont {Bonneau}},
  \bibinfo {author} {\bibfnamefont {R.}~\bibnamefont {Lopes}}, \bibinfo
  {author} {\bibfnamefont {J.}~\bibnamefont {Ruaudel}}, \bibinfo {author}
  {\bibfnamefont {D.}~\bibnamefont {Boiron}},\ and\ \bibinfo {author}
  {\bibfnamefont {C.~I.}\ \bibnamefont {Westbrook}},\ }\bibfield  {title}
  {\bibinfo {title} {{Acoustic Analog to the Dynamical Casimir Effect in a
  Bose-Einstein Condensate}},\ }\href
  {https://doi.org/10.1103/PhysRevLett.109.220401} {\bibfield  {journal}
  {\bibinfo  {journal} {Phys. Rev. Lett.}\ }\textbf {\bibinfo {volume} {109}},\
  \bibinfo {pages} {220401} (\bibinfo {year} {2012})}\BibitemShut {NoStop}%
\bibitem [{\citenamefont {Recati}\ \emph {et~al.}(2009)\citenamefont {Recati},
  \citenamefont {Pavloff},\ and\ \citenamefont
  {Carusotto}}]{PhysRevA.80.043603}%
  \BibitemOpen
  \bibfield  {author} {\bibinfo {author} {\bibfnamefont {A.}~\bibnamefont
  {Recati}}, \bibinfo {author} {\bibfnamefont {N.}~\bibnamefont {Pavloff}},\
  and\ \bibinfo {author} {\bibfnamefont {I.}~\bibnamefont {Carusotto}},\
  }\bibfield  {title} {\bibinfo {title} {{Bogoliubov theory of acoustic Hawking
  radiation in Bose-Einstein condensates}},\ }\href
  {https://doi.org/10.1103/PhysRevA.80.043603} {\bibfield  {journal} {\bibinfo
  {journal} {Phys. Rev. A}\ }\textbf {\bibinfo {volume} {80}},\ \bibinfo
  {pages} {043603} (\bibinfo {year} {2009})}\BibitemShut {NoStop}%
\bibitem [{\citenamefont {Larr\'e}\ \emph {et~al.}(2012)\citenamefont
  {Larr\'e}, \citenamefont {Recati}, \citenamefont {Carusotto},\ and\
  \citenamefont {Pavloff}}]{PhysRevA.85.013621}%
  \BibitemOpen
  \bibfield  {author} {\bibinfo {author} {\bibfnamefont {P.-E.}\ \bibnamefont
  {Larr\'e}}, \bibinfo {author} {\bibfnamefont {A.}~\bibnamefont {Recati}},
  \bibinfo {author} {\bibfnamefont {I.}~\bibnamefont {Carusotto}},\ and\
  \bibinfo {author} {\bibfnamefont {N.}~\bibnamefont {Pavloff}},\ }\bibfield
  {title} {\bibinfo {title} {{Quantum fluctuations around black hole horizons
  in Bose-Einstein condensates}},\ }\href
  {https://doi.org/10.1103/PhysRevA.85.013621} {\bibfield  {journal} {\bibinfo
  {journal} {Phys. Rev. A}\ }\textbf {\bibinfo {volume} {85}},\ \bibinfo
  {pages} {013621} (\bibinfo {year} {2012})}\BibitemShut {NoStop}%
\bibitem [{\citenamefont {Coutant}\ and\ \citenamefont
  {Weinfurtner}(2018)}]{PhysRevD.97.025006}%
  \BibitemOpen
  \bibfield  {author} {\bibinfo {author} {\bibfnamefont {A.}~\bibnamefont
  {Coutant}}\ and\ \bibinfo {author} {\bibfnamefont {S.}~\bibnamefont
  {Weinfurtner}},\ }\bibfield  {title} {\bibinfo {title} {{Low-frequency
  analogue Hawking radiation: The Bogoliubov-de Gennes model}},\ }\href
  {https://doi.org/10.1103/PhysRevD.97.025006} {\bibfield  {journal} {\bibinfo
  {journal} {Phys. Rev. D}\ }\textbf {\bibinfo {volume} {97}},\ \bibinfo
  {pages} {025006} (\bibinfo {year} {2018})}\BibitemShut {NoStop}%
\bibitem [{\citenamefont {Bordag}\ \emph {et~al.}(1999)\citenamefont {Bordag},
  \citenamefont {Kirsten},\ and\ \citenamefont
  {Vassilevich}}]{PhysRevD.59.085011}%
  \BibitemOpen
  \bibfield  {author} {\bibinfo {author} {\bibfnamefont {M.}~\bibnamefont
  {Bordag}}, \bibinfo {author} {\bibfnamefont {K.}~\bibnamefont {Kirsten}},\
  and\ \bibinfo {author} {\bibfnamefont {D.}~\bibnamefont {Vassilevich}},\
  }\bibfield  {title} {\bibinfo {title} {{Ground state energy for a penetrable
  sphere and for a dielectric ball}},\ }\href
  {https://doi.org/10.1103/PhysRevD.59.085011} {\bibfield  {journal} {\bibinfo
  {journal} {Phys. Rev. D}\ }\textbf {\bibinfo {volume} {59}},\ \bibinfo
  {pages} {085011} (\bibinfo {year} {1999})}\BibitemShut {NoStop}%
\bibitem [{\citenamefont {Hawking}(1974)}]{HAWKING1974}%
  \BibitemOpen
  \bibfield  {author} {\bibinfo {author} {\bibfnamefont {S.~W.}\ \bibnamefont
  {Hawking}},\ }\bibfield  {title} {\bibinfo {title} {{Black hole
  explosions?}},\ }\href {https://doi.org/10.1038/248030a0} {\bibfield
  {journal} {\bibinfo  {journal} {Nature}\ }\textbf {\bibinfo {volume} {248}},\
  \bibinfo {pages} {30} (\bibinfo {year} {1974})}\BibitemShut {NoStop}%
\bibitem [{\citenamefont {Cheek}\ \emph {et~al.}(2023)\citenamefont {Cheek},
  \citenamefont {Heurtier}, \citenamefont {Perez-Gonzalez},\ and\ \citenamefont
  {Turner}}]{Cheek}%
  \BibitemOpen
  \bibfield  {author} {\bibinfo {author} {\bibfnamefont {A.}~\bibnamefont
  {Cheek}}, \bibinfo {author} {\bibfnamefont {L.}~\bibnamefont {Heurtier}},
  \bibinfo {author} {\bibfnamefont {Y.~F.}\ \bibnamefont {Perez-Gonzalez}},\
  and\ \bibinfo {author} {\bibfnamefont {J.}~\bibnamefont {Turner}},\
  }\bibfield  {title} {\bibinfo {title} {{Evaporation of primordial black holes
  in the early Universe: Mass and spin distributions}},\ }\href
  {https://doi.org/10.1103/PhysRevD.108.015005} {\bibfield  {journal} {\bibinfo
   {journal} {Phys. Rev. D}\ }\textbf {\bibinfo {volume} {108}},\ \bibinfo
  {pages} {015005} (\bibinfo {year} {2023})}\BibitemShut {NoStop}%
\bibitem [{\citenamefont {Auffinger}(2023)}]{AUFFINGER2023104040}%
  \BibitemOpen
  \bibfield  {author} {\bibinfo {author} {\bibfnamefont {J.}~\bibnamefont
  {Auffinger}},\ }\bibfield  {title} {\bibinfo {title} {{Primordial black hole
  constraints with Hawking radiation—A review}},\ }\href
  {https://doi.org/https://doi.org/10.1016/j.ppnp.2023.104040} {\bibfield
  {journal} {\bibinfo  {journal} {Progress in Particle and Nuclear Physics}\
  }\textbf {\bibinfo {volume} {131}},\ \bibinfo {pages} {104040} (\bibinfo
  {year} {2023})}\BibitemShut {NoStop}%
\bibitem [{\citenamefont {Balbinot}\ \emph {et~al.}(2005)\citenamefont
  {Balbinot}, \citenamefont {Fagnocchi}, \citenamefont {Fabbri},\ and\
  \citenamefont {Procopio}}]{PhysRevLett.94.161302}%
  \BibitemOpen
  \bibfield  {author} {\bibinfo {author} {\bibfnamefont {R.}~\bibnamefont
  {Balbinot}}, \bibinfo {author} {\bibfnamefont {S.}~\bibnamefont {Fagnocchi}},
  \bibinfo {author} {\bibfnamefont {A.}~\bibnamefont {Fabbri}},\ and\ \bibinfo
  {author} {\bibfnamefont {G.~P.}\ \bibnamefont {Procopio}},\ }\bibfield
  {title} {\bibinfo {title} {{Backreaction in Acoustic Black Holes}},\ }\href
  {https://doi.org/10.1103/PhysRevLett.94.161302} {\bibfield  {journal}
  {\bibinfo  {journal} {Phys. Rev. Lett.}\ }\textbf {\bibinfo {volume} {94}},\
  \bibinfo {pages} {161302} (\bibinfo {year} {2005})}\BibitemShut {NoStop}%
\bibitem [{\citenamefont {Sch\"utzhold}\ \emph {et~al.}(2005)\citenamefont
  {Sch\"utzhold}, \citenamefont {Uhlmann}, \citenamefont {Xu},\ and\
  \citenamefont {Fischer}}]{PhysRevD.72.105005}%
  \BibitemOpen
  \bibfield  {author} {\bibinfo {author} {\bibfnamefont {R.}~\bibnamefont
  {Sch\"utzhold}}, \bibinfo {author} {\bibfnamefont {M.}~\bibnamefont
  {Uhlmann}}, \bibinfo {author} {\bibfnamefont {Y.}~\bibnamefont {Xu}},\ and\
  \bibinfo {author} {\bibfnamefont {U.~R.}\ \bibnamefont {Fischer}},\
  }\bibfield  {title} {\bibinfo {title} {{Quantum backreaction in dilute
  Bose-Einstein condensates}},\ }\href
  {https://doi.org/10.1103/PhysRevD.72.105005} {\bibfield  {journal} {\bibinfo
  {journal} {Phys. Rev. D}\ }\textbf {\bibinfo {volume} {72}},\ \bibinfo
  {pages} {105005} (\bibinfo {year} {2005})}\BibitemShut {NoStop}%
\bibitem [{\citenamefont {Liberati}\ \emph {et~al.}(2020)\citenamefont
  {Liberati}, \citenamefont {Tricella},\ and\ \citenamefont
  {Trombettoni}}]{Liberati2020}%
  \BibitemOpen
  \bibfield  {author} {\bibinfo {author} {\bibfnamefont {S.}~\bibnamefont
  {Liberati}}, \bibinfo {author} {\bibfnamefont {G.}~\bibnamefont {Tricella}},\
  and\ \bibinfo {author} {\bibfnamefont {A.}~\bibnamefont {Trombettoni}},\
  }\bibfield  {title} {\bibinfo {title} {{Back-Reaction in Canonical Analogue
  Black Holes}},\ }\bibfield  {journal} {\bibinfo  {journal} {Applied
  Sciences}\ }\textbf {\bibinfo {volume} {10}},\ \href
  {https://doi.org/10.3390/app10248868} {10.3390/app10248868} (\bibinfo {year}
  {2020})\BibitemShut {NoStop}%
\bibitem [{\citenamefont {Baak}\ \emph {et~al.}(2022)\citenamefont {Baak},
  \citenamefont {Ribeiro},\ and\ \citenamefont {Fischer}}]{RibeiroPRA}%
  \BibitemOpen
  \bibfield  {author} {\bibinfo {author} {\bibfnamefont {S.-S.}\ \bibnamefont
  {Baak}}, \bibinfo {author} {\bibfnamefont {C.~C.~H.}\ \bibnamefont
  {Ribeiro}},\ and\ \bibinfo {author} {\bibfnamefont {U.~R.}\ \bibnamefont
  {Fischer}},\ }\bibfield  {title} {\bibinfo {title} {{Number-conserving
  solution for dynamical quantum backreaction in a Bose-Einstein condensate}},\
  }\href {https://doi.org/10.1103/PhysRevA.106.053319} {\bibfield  {journal}
  {\bibinfo  {journal} {Phys. Rev. A}\ }\textbf {\bibinfo {volume} {106}},\
  \bibinfo {pages} {053319} (\bibinfo {year} {2022})}\BibitemShut {NoStop}%
\bibitem [{\citenamefont {Butera}\ and\ \citenamefont
  {Carusotto}(2023)}]{PhysRevLett.130.241501}%
  \BibitemOpen
  \bibfield  {author} {\bibinfo {author} {\bibfnamefont {S.}~\bibnamefont
  {Butera}}\ and\ \bibinfo {author} {\bibfnamefont {I.}~\bibnamefont
  {Carusotto}},\ }\bibfield  {title} {\bibinfo {title} {{Numerical Studies of
  Back Reaction Effects in an Analog Model of Cosmological Preheating}},\
  }\href {https://doi.org/10.1103/PhysRevLett.130.241501} {\bibfield  {journal}
  {\bibinfo  {journal} {Phys. Rev. Lett.}\ }\textbf {\bibinfo {volume} {130}},\
  \bibinfo {pages} {241501} (\bibinfo {year} {2023})}\BibitemShut {NoStop}%
\bibitem [{\citenamefont {Weinfurtner}(2005)}]{Weinfurtner2005}%
  \BibitemOpen
  \bibfield  {author} {\bibinfo {author} {\bibfnamefont {S.}~\bibnamefont
  {Weinfurtner}},\ }\bibfield  {title} {\bibinfo {title} {{Analogue model for
  an expanding universe}},\ }\href {https://doi.org/10.1007/s10714-005-0135-7}
  {\bibfield  {journal} {\bibinfo  {journal} {General Relativity and
  Gravitation}\ }\textbf {\bibinfo {volume} {37}},\ \bibinfo {pages} {1549}
  (\bibinfo {year} {2005})}\BibitemShut {NoStop}%
\bibitem [{\citenamefont {Ch\"a}\ and\ \citenamefont
  {Fischer}(2017)}]{PhysRevLett.118.130404}%
  \BibitemOpen
  \bibfield  {author} {\bibinfo {author} {\bibfnamefont {S.-Y.}\ \bibnamefont
  {Ch\"a}}\ and\ \bibinfo {author} {\bibfnamefont {U.~R.}\ \bibnamefont
  {Fischer}},\ }\bibfield  {title} {\bibinfo {title} {{Probing the Scale
  Invariance of the Inflationary Power Spectrum in Expanding
  Quasi-Two-Dimensional Dipolar Condensates}},\ }\href
  {https://doi.org/10.1103/PhysRevLett.118.130404} {\bibfield  {journal}
  {\bibinfo  {journal} {Phys. Rev. Lett.}\ }\textbf {\bibinfo {volume} {118}},\
  \bibinfo {pages} {130404} (\bibinfo {year} {2017})}\BibitemShut {NoStop}%
\bibitem [{\citenamefont {Jain}\ \emph {et~al.}(2007)\citenamefont {Jain},
  \citenamefont {Weinfurtner}, \citenamefont {Visser},\ and\ \citenamefont
  {Gardiner}}]{Jain}%
  \BibitemOpen
  \bibfield  {author} {\bibinfo {author} {\bibfnamefont {P.}~\bibnamefont
  {Jain}}, \bibinfo {author} {\bibfnamefont {S.}~\bibnamefont {Weinfurtner}},
  \bibinfo {author} {\bibfnamefont {M.}~\bibnamefont {Visser}},\ and\ \bibinfo
  {author} {\bibfnamefont {C.~W.}\ \bibnamefont {Gardiner}},\ }\bibfield
  {title} {\bibinfo {title} {{Analog model of a Friedmann-Robertson-Walker
  universe in Bose-Einstein condensates: Application of the classical field
  method}},\ }\href {https://doi.org/10.1103/PhysRevA.76.033616} {\bibfield
  {journal} {\bibinfo  {journal} {Phys. Rev. A}\ }\textbf {\bibinfo {volume}
  {76}},\ \bibinfo {pages} {033616} (\bibinfo {year} {2007})}\BibitemShut
  {NoStop}%
\bibitem [{\citenamefont {Zache}\ \emph {et~al.}(2017)\citenamefont {Zache},
  \citenamefont {Kasper},\ and\ \citenamefont {Berges}}]{zache2017}%
  \BibitemOpen
  \bibfield  {author} {\bibinfo {author} {\bibfnamefont {T.~V.}\ \bibnamefont
  {Zache}}, \bibinfo {author} {\bibfnamefont {V.}~\bibnamefont {Kasper}},\ and\
  \bibinfo {author} {\bibfnamefont {J.}~\bibnamefont {Berges}},\ }\bibfield
  {title} {\bibinfo {title} {{Inflationary preheating dynamics with two-species
  condensates}},\ }\href {https://doi.org/10.1103/PhysRevA.95.063629}
  {\bibfield  {journal} {\bibinfo  {journal} {Phys. Rev. A}\ }\textbf {\bibinfo
  {volume} {95}},\ \bibinfo {pages} {063629} (\bibinfo {year}
  {2017})}\BibitemShut {NoStop}%
\bibitem [{\citenamefont {Gomez~Llorente}\ and\ \citenamefont
  {Plata}(2019)}]{PhysRevA.100.043613}%
  \BibitemOpen
  \bibfield  {author} {\bibinfo {author} {\bibfnamefont {J.~M.}\ \bibnamefont
  {Gomez~Llorente}}\ and\ \bibinfo {author} {\bibfnamefont {J.}~\bibnamefont
  {Plata}},\ }\bibfield  {title} {\bibinfo {title} {{Expanding ring-shaped
  Bose-Einstein condensates as analogs of cosmological models: Analytical
  characterization of the inflationary dynamics}},\ }\href
  {https://doi.org/10.1103/PhysRevA.100.043613} {\bibfield  {journal} {\bibinfo
   {journal} {Phys. Rev. A}\ }\textbf {\bibinfo {volume} {100}},\ \bibinfo
  {pages} {043613} (\bibinfo {year} {2019})}\BibitemShut {NoStop}%
\bibitem [{\citenamefont {Chatrchyan}\ \emph {et~al.}(2021)\citenamefont
  {Chatrchyan}, \citenamefont {Geier}, \citenamefont {Oberthaler},
  \citenamefont {Berges},\ and\ \citenamefont {Hauke}}]{PhysRevA.104.023302}%
  \BibitemOpen
  \bibfield  {author} {\bibinfo {author} {\bibfnamefont {A.}~\bibnamefont
  {Chatrchyan}}, \bibinfo {author} {\bibfnamefont {K.~T.}\ \bibnamefont
  {Geier}}, \bibinfo {author} {\bibfnamefont {M.~K.}\ \bibnamefont
  {Oberthaler}}, \bibinfo {author} {\bibfnamefont {J.}~\bibnamefont {Berges}},\
  and\ \bibinfo {author} {\bibfnamefont {P.}~\bibnamefont {Hauke}},\ }\bibfield
   {title} {\bibinfo {title} {{Analog cosmological reheating in an ultracold
  Bose gas}},\ }\href {https://doi.org/10.1103/PhysRevA.104.023302} {\bibfield
  {journal} {\bibinfo  {journal} {Phys. Rev. A}\ }\textbf {\bibinfo {volume}
  {104}},\ \bibinfo {pages} {023302} (\bibinfo {year} {2021})}\BibitemShut
  {NoStop}%
\bibitem [{\citenamefont {Tolosa-Sime\'on}\ \emph {et~al.}(2022)\citenamefont
  {Tolosa-Sime\'on}, \citenamefont {Parra-L\'opez}, \citenamefont
  {S\'anchez-Kuntz}, \citenamefont {Haas}, \citenamefont {Viermann},
  \citenamefont {Sparn}, \citenamefont {Liebster}, \citenamefont {Hans},
  \citenamefont {Kath}, \citenamefont {Strobel}, \citenamefont {Oberthaler},\
  and\ \citenamefont {Floerchinger}}]{Mireia}%
  \BibitemOpen
  \bibfield  {author} {\bibinfo {author} {\bibfnamefont {M.}~\bibnamefont
  {Tolosa-Sime\'on}}, \bibinfo {author} {\bibfnamefont {A.}~\bibnamefont
  {Parra-L\'opez}}, \bibinfo {author} {\bibfnamefont {N.}~\bibnamefont
  {S\'anchez-Kuntz}}, \bibinfo {author} {\bibfnamefont {T.}~\bibnamefont
  {Haas}}, \bibinfo {author} {\bibfnamefont {C.}~\bibnamefont {Viermann}},
  \bibinfo {author} {\bibfnamefont {M.}~\bibnamefont {Sparn}}, \bibinfo
  {author} {\bibfnamefont {N.}~\bibnamefont {Liebster}}, \bibinfo {author}
  {\bibfnamefont {M.}~\bibnamefont {Hans}}, \bibinfo {author} {\bibfnamefont
  {E.}~\bibnamefont {Kath}}, \bibinfo {author} {\bibfnamefont {H.}~\bibnamefont
  {Strobel}}, \bibinfo {author} {\bibfnamefont {M.~K.}\ \bibnamefont
  {Oberthaler}},\ and\ \bibinfo {author} {\bibfnamefont {S.}~\bibnamefont
  {Floerchinger}},\ }\bibfield  {title} {\bibinfo {title} {{Curved and
  expanding spacetime geometries in Bose-Einstein condensates}},\ }\href
  {https://doi.org/10.1103/PhysRevA.106.033313} {\bibfield  {journal} {\bibinfo
   {journal} {Phys. Rev. A}\ }\textbf {\bibinfo {volume} {106}},\ \bibinfo
  {pages} {033313} (\bibinfo {year} {2022})}\BibitemShut {NoStop}%
\bibitem [{\citenamefont {Bhardwaj}\ \emph {et~al.}(2021)\citenamefont
  {Bhardwaj}, \citenamefont {Vaido},\ and\ \citenamefont
  {Sheehy}}]{PhysRevA.103.023322}%
  \BibitemOpen
  \bibfield  {author} {\bibinfo {author} {\bibfnamefont {A.}~\bibnamefont
  {Bhardwaj}}, \bibinfo {author} {\bibfnamefont {D.}~\bibnamefont {Vaido}},\
  and\ \bibinfo {author} {\bibfnamefont {D.~E.}\ \bibnamefont {Sheehy}},\
  }\bibfield  {title} {\bibinfo {title} {{Inflationary dynamics and particle
  production in a toroidal Bose-Einstein condensate}},\ }\href
  {https://doi.org/10.1103/PhysRevA.103.023322} {\bibfield  {journal} {\bibinfo
   {journal} {Phys. Rev. A}\ }\textbf {\bibinfo {volume} {103}},\ \bibinfo
  {pages} {023322} (\bibinfo {year} {2021})}\BibitemShut {NoStop}%
\bibitem [{\citenamefont {Schmidt}\ \emph {et~al.}(2024)\citenamefont
  {Schmidt}, \citenamefont {Parra-L\'opez}, \citenamefont {Tolosa-Sime\'on},
  \citenamefont {Sparn}, \citenamefont {Kath}, \citenamefont {Liebster},
  \citenamefont {Duchene}, \citenamefont {Strobel}, \citenamefont
  {Oberthaler},\ and\ \citenamefont {Floerchinger}}]{PhysRevD.110.123523}%
  \BibitemOpen
  \bibfield  {author} {\bibinfo {author} {\bibfnamefont {C.~F.}\ \bibnamefont
  {Schmidt}}, \bibinfo {author} {\bibfnamefont {A.}~\bibnamefont
  {Parra-L\'opez}}, \bibinfo {author} {\bibfnamefont {M.}~\bibnamefont
  {Tolosa-Sime\'on}}, \bibinfo {author} {\bibfnamefont {M.}~\bibnamefont
  {Sparn}}, \bibinfo {author} {\bibfnamefont {E.}~\bibnamefont {Kath}},
  \bibinfo {author} {\bibfnamefont {N.}~\bibnamefont {Liebster}}, \bibinfo
  {author} {\bibfnamefont {J.}~\bibnamefont {Duchene}}, \bibinfo {author}
  {\bibfnamefont {H.}~\bibnamefont {Strobel}}, \bibinfo {author} {\bibfnamefont
  {M.~K.}\ \bibnamefont {Oberthaler}},\ and\ \bibinfo {author} {\bibfnamefont
  {S.}~\bibnamefont {Floerchinger}},\ }\bibfield  {title} {\bibinfo {title}
  {{Cosmological particle production in a quantum field simulator as a quantum
  mechanical scattering problem}},\ }\href
  {https://doi.org/10.1103/PhysRevD.110.123523} {\bibfield  {journal} {\bibinfo
   {journal} {Phys. Rev. D}\ }\textbf {\bibinfo {volume} {110}},\ \bibinfo
  {pages} {123523} (\bibinfo {year} {2024})}\BibitemShut {NoStop}%
\bibitem [{\citenamefont {Béssa}\ \emph {et~al.}(2009)\citenamefont {Béssa},
  \citenamefont {Bezerra},\ and\ \citenamefont {Ford}}]{bessa}%
  \BibitemOpen
  \bibfield  {author} {\bibinfo {author} {\bibfnamefont {C.~H.~G.}\
  \bibnamefont {Béssa}}, \bibinfo {author} {\bibfnamefont {V.~B.}\
  \bibnamefont {Bezerra}},\ and\ \bibinfo {author} {\bibfnamefont {L.~H.}\
  \bibnamefont {Ford}},\ }\bibfield  {title} {\bibinfo {title} {{Brownian
  motion in Robertson–Walker spacetimes from electromagnetic vacuum
  fluctuations}},\ }\href {https://doi.org/10.1063/1.3133946} {\bibfield
  {journal} {\bibinfo  {journal} {Journal of Mathematical Physics}\ }\textbf
  {\bibinfo {volume} {50}},\ \bibinfo {pages} {062501} (\bibinfo {year}
  {2009})},\ \Eprint
  {https://arxiv.org/abs/https://pubs.aip.org/aip/jmp/article-pdf/doi/10.1063/1.3133946/15795955/062501\_1\_online.pdf}
  {https://pubs.aip.org/aip/jmp/article-pdf/doi/10.1063/1.3133946/15795955/062501\_1\_online.pdf}
  \BibitemShut {NoStop}%
\bibitem [{\citenamefont {Li}\ \emph {et~al.}(2010)\citenamefont {Li},
  \citenamefont {Wu},\ and\ \citenamefont {Yu}}]{Zhengxiang}%
  \BibitemOpen
  \bibfield  {author} {\bibinfo {author} {\bibfnamefont {Z.}~\bibnamefont
  {Li}}, \bibinfo {author} {\bibfnamefont {P.}~\bibnamefont {Wu}},\ and\
  \bibinfo {author} {\bibfnamefont {H.}~\bibnamefont {Yu}},\ }\bibfield
  {title} {\bibinfo {title} {{Probing the course of cosmic expansion with a
  combination of observational data}},\ }\href
  {https://doi.org/10.1088/1475-7516/2010/11/031} {\bibfield  {journal}
  {\bibinfo  {journal} {Journal of Cosmology and Astroparticle Physics}\
  }\textbf {\bibinfo {volume} {2010}}\bibinfo  {number} { (11)},\ \bibinfo
  {pages} {031}}\BibitemShut {NoStop}%
\bibitem [{\citenamefont {Huterer}\ \emph {et~al.}(2015)\citenamefont
  {Huterer}, \citenamefont {Kirkby}, \citenamefont {Bean}, \citenamefont
  {Connolly}, \citenamefont {Dawson}, \citenamefont {Dodelson}, \citenamefont
  {Evrard}, \citenamefont {Jain}, \citenamefont {Jarvis}, \citenamefont
  {Linder}, \citenamefont {Mandelbaum}, \citenamefont {May}, \citenamefont
  {Raccanelli}, \citenamefont {Reid}, \citenamefont {Rozo}, \citenamefont
  {Schmidt}, \citenamefont {Sehgal}, \citenamefont {Slosar}, \citenamefont
  {{van Engelen}}, \citenamefont {Wu},\ and\ \citenamefont
  {Zhao}}]{HUTERER201523}%
  \BibitemOpen
\bibfield  {number} {  }\bibfield  {author} {\bibinfo {author} {\bibfnamefont
  {D.}~\bibnamefont {Huterer}}, \bibinfo {author} {\bibfnamefont
  {D.}~\bibnamefont {Kirkby}}, \bibinfo {author} {\bibfnamefont
  {R.}~\bibnamefont {Bean}}, \bibinfo {author} {\bibfnamefont {A.}~\bibnamefont
  {Connolly}}, \bibinfo {author} {\bibfnamefont {K.}~\bibnamefont {Dawson}},
  \bibinfo {author} {\bibfnamefont {S.}~\bibnamefont {Dodelson}}, \bibinfo
  {author} {\bibfnamefont {A.}~\bibnamefont {Evrard}}, \bibinfo {author}
  {\bibfnamefont {B.}~\bibnamefont {Jain}}, \bibinfo {author} {\bibfnamefont
  {M.}~\bibnamefont {Jarvis}}, \bibinfo {author} {\bibfnamefont
  {E.}~\bibnamefont {Linder}}, \bibinfo {author} {\bibfnamefont
  {R.}~\bibnamefont {Mandelbaum}}, \bibinfo {author} {\bibfnamefont
  {M.}~\bibnamefont {May}}, \bibinfo {author} {\bibfnamefont {A.}~\bibnamefont
  {Raccanelli}}, \bibinfo {author} {\bibfnamefont {B.}~\bibnamefont {Reid}},
  \bibinfo {author} {\bibfnamefont {E.}~\bibnamefont {Rozo}}, \bibinfo {author}
  {\bibfnamefont {F.}~\bibnamefont {Schmidt}}, \bibinfo {author} {\bibfnamefont
  {N.}~\bibnamefont {Sehgal}}, \bibinfo {author} {\bibfnamefont
  {A.}~\bibnamefont {Slosar}}, \bibinfo {author} {\bibfnamefont
  {A.}~\bibnamefont {{van Engelen}}}, \bibinfo {author} {\bibfnamefont {H.-Y.}\
  \bibnamefont {Wu}},\ and\ \bibinfo {author} {\bibfnamefont {G.}~\bibnamefont
  {Zhao}},\ }\bibfield  {title} {\bibinfo {title} {{Growth of cosmic structure:
  Probing dark energy beyond expansion}},\ }\href
  {https://doi.org/https://doi.org/10.1016/j.astropartphys.2014.07.004}
  {\bibfield  {journal} {\bibinfo  {journal} {Astroparticle Physics}\ }\textbf
  {\bibinfo {volume} {63}},\ \bibinfo {pages} {23} (\bibinfo {year} {2015})},\
  \bibinfo {note} {dark Energy and CMB}\BibitemShut {NoStop}%
\bibitem [{\citenamefont {Moresco}\ \emph {et~al.}(2022)\citenamefont
  {Moresco}, \citenamefont {Amati}, \citenamefont {Amendola}, \citenamefont
  {Birrer}, \citenamefont {Blakeslee}, \citenamefont {Cantiello}, \citenamefont
  {Cimatti}, \citenamefont {Darling}, \citenamefont {Della~Valle},
  \citenamefont {Fishbach}, \citenamefont {Grillo}, \citenamefont {Hamaus},
  \citenamefont {Holz}, \citenamefont {Izzo}, \citenamefont {Jimenez},
  \citenamefont {Lusso}, \citenamefont {Meneghetti}, \citenamefont
  {Piedipalumbo}, \citenamefont {Pisani}, \citenamefont {Pourtsidou},
  \citenamefont {Pozzetti}, \citenamefont {Quartin}, \citenamefont {Risaliti},
  \citenamefont {Rosati},\ and\ \citenamefont {Verde}}]{Moresco2022}%
  \BibitemOpen
  \bibfield  {author} {\bibinfo {author} {\bibfnamefont {M.}~\bibnamefont
  {Moresco}}, \bibinfo {author} {\bibfnamefont {L.}~\bibnamefont {Amati}},
  \bibinfo {author} {\bibfnamefont {L.}~\bibnamefont {Amendola}}, \bibinfo
  {author} {\bibfnamefont {S.}~\bibnamefont {Birrer}}, \bibinfo {author}
  {\bibfnamefont {J.~P.}\ \bibnamefont {Blakeslee}}, \bibinfo {author}
  {\bibfnamefont {M.}~\bibnamefont {Cantiello}}, \bibinfo {author}
  {\bibfnamefont {A.}~\bibnamefont {Cimatti}}, \bibinfo {author} {\bibfnamefont
  {J.}~\bibnamefont {Darling}}, \bibinfo {author} {\bibfnamefont
  {M.}~\bibnamefont {Della~Valle}}, \bibinfo {author} {\bibfnamefont
  {M.}~\bibnamefont {Fishbach}}, \bibinfo {author} {\bibfnamefont
  {C.}~\bibnamefont {Grillo}}, \bibinfo {author} {\bibfnamefont
  {N.}~\bibnamefont {Hamaus}}, \bibinfo {author} {\bibfnamefont
  {D.}~\bibnamefont {Holz}}, \bibinfo {author} {\bibfnamefont {L.}~\bibnamefont
  {Izzo}}, \bibinfo {author} {\bibfnamefont {R.}~\bibnamefont {Jimenez}},
  \bibinfo {author} {\bibfnamefont {E.}~\bibnamefont {Lusso}}, \bibinfo
  {author} {\bibfnamefont {M.}~\bibnamefont {Meneghetti}}, \bibinfo {author}
  {\bibfnamefont {E.}~\bibnamefont {Piedipalumbo}}, \bibinfo {author}
  {\bibfnamefont {A.}~\bibnamefont {Pisani}}, \bibinfo {author} {\bibfnamefont
  {A.}~\bibnamefont {Pourtsidou}}, \bibinfo {author} {\bibfnamefont
  {L.}~\bibnamefont {Pozzetti}}, \bibinfo {author} {\bibfnamefont
  {M.}~\bibnamefont {Quartin}}, \bibinfo {author} {\bibfnamefont
  {G.}~\bibnamefont {Risaliti}}, \bibinfo {author} {\bibfnamefont
  {P.}~\bibnamefont {Rosati}},\ and\ \bibinfo {author} {\bibfnamefont
  {L.}~\bibnamefont {Verde}},\ }\bibfield  {title} {\bibinfo {title}
  {{Unveiling the Universe with emerging cosmological probes}},\ }\href
  {https://doi.org/10.1007/s41114-022-00040-z} {\bibfield  {journal} {\bibinfo
  {journal} {Living Reviews in Relativity}\ }\textbf {\bibinfo {volume} {25}},\
  \bibinfo {pages} {6} (\bibinfo {year} {2022})}\BibitemShut {NoStop}%
\bibitem [{\citenamefont {Penna-Lima}\ \emph {et~al.}(2023)\citenamefont
  {Penna-Lima}, \citenamefont {Pinto-Neto},\ and\ \citenamefont
  {Vitenti}}]{Sandro}%
  \BibitemOpen
  \bibfield  {author} {\bibinfo {author} {\bibfnamefont {M.}~\bibnamefont
  {Penna-Lima}}, \bibinfo {author} {\bibfnamefont {N.}~\bibnamefont
  {Pinto-Neto}},\ and\ \bibinfo {author} {\bibfnamefont {S.~D.~P.}\
  \bibnamefont {Vitenti}},\ }\bibfield  {title} {\bibinfo {title} {{New
  formalism to define vacuum states for scalar fields in curved spacetimes}},\
  }\href {https://doi.org/10.1103/PhysRevD.107.065019} {\bibfield  {journal}
  {\bibinfo  {journal} {Phys. Rev. D}\ }\textbf {\bibinfo {volume} {107}},\
  \bibinfo {pages} {065019} (\bibinfo {year} {2023})}\BibitemShut {NoStop}%
\bibitem [{\citenamefont {Birrell}\ and\ \citenamefont
  {Davies}(1982)}]{Birrell}%
  \BibitemOpen
  \bibfield  {author} {\bibinfo {author} {\bibfnamefont {N.~D.}\ \bibnamefont
  {Birrell}}\ and\ \bibinfo {author} {\bibfnamefont {P.~C.~W.}\ \bibnamefont
  {Davies}},\ }\href {https://doi.org/10.1017/CBO9780511622632} {\emph
  {\bibinfo {title} {{Quantum Fields in Curved Space}}}},\ Cambridge Monographs
  on Mathematical Physics\ (\bibinfo  {publisher} {Cambridge University
  Press},\ \bibinfo {address} {Cambridge, UK},\ \bibinfo {year}
  {1982})\BibitemShut {NoStop}%
\bibitem [{\citenamefont {Navon}\ \emph {et~al.}(2021)\citenamefont {Navon},
  \citenamefont {Smith},\ and\ \citenamefont {Hadzibabic}}]{Navon2021}%
  \BibitemOpen
  \bibfield  {author} {\bibinfo {author} {\bibfnamefont {N.}~\bibnamefont
  {Navon}}, \bibinfo {author} {\bibfnamefont {R.~P.}\ \bibnamefont {Smith}},\
  and\ \bibinfo {author} {\bibfnamefont {Z.}~\bibnamefont {Hadzibabic}},\
  }\bibfield  {title} {\bibinfo {title} {{Quantum gases in optical boxes}},\
  }\href {https://doi.org/10.1038/s41567-021-01403-z} {\bibfield  {journal}
  {\bibinfo  {journal} {Nature Physics}\ }\textbf {\bibinfo {volume} {17}},\
  \bibinfo {pages} {1334} (\bibinfo {year} {2021})}\BibitemShut {NoStop}%
\bibitem [{sup()}]{supp}%
  \BibitemOpen
  \href@noop {} {}\bibinfo {note} {See Supplemental Material for a de tailed
  discussion and derivations.}\BibitemShut {Stop}%
\bibitem [{\citenamefont {Castin}\ and\ \citenamefont {Dum}(1998)}]{castin}%
  \BibitemOpen
  \bibfield  {author} {\bibinfo {author} {\bibfnamefont {Y.}~\bibnamefont
  {Castin}}\ and\ \bibinfo {author} {\bibfnamefont {R.}~\bibnamefont {Dum}},\
  }\bibfield  {title} {\bibinfo {title} {{Low-temperature Bose-Einstein
  condensates in time-dependent traps: Beyond the $U(1)$ symmetry-breaking
  approach}},\ }\href {https://doi.org/10.1103/PhysRevA.57.3008} {\bibfield
  {journal} {\bibinfo  {journal} {Phys. Rev. A}\ }\textbf {\bibinfo {volume}
  {57}},\ \bibinfo {pages} {3008} (\bibinfo {year} {1998})}\BibitemShut
  {NoStop}%
\bibitem [{\citenamefont {Holanda~Ribeiro}(2025)}]{PhysRevA.111.023306}%
  \BibitemOpen
  \bibfield  {author} {\bibinfo {author} {\bibfnamefont {C.~C.}\ \bibnamefont
  {Holanda~Ribeiro}},\ }\bibfield  {title} {\bibinfo {title} {{Energy
  conservation and quantum backreaction in Bose-Einstein condensates}},\ }\href
  {https://doi.org/10.1103/PhysRevA.111.023306} {\bibfield  {journal} {\bibinfo
   {journal} {Phys. Rev. A}\ }\textbf {\bibinfo {volume} {111}},\ \bibinfo
  {pages} {023306} (\bibinfo {year} {2025})}\BibitemShut {NoStop}%
\bibitem [{Note1()}]{Note1}%
  \BibitemOpen
  \bibinfo {note} {Usually the density variance is denoted by
  $G^{(2)}$}\BibitemShut {NoStop}%
\bibitem [{\citenamefont {Lewenstein}\ and\ \citenamefont {You}(1996)}]{lew}%
  \BibitemOpen
  \bibfield  {author} {\bibinfo {author} {\bibfnamefont {M.}~\bibnamefont
  {Lewenstein}}\ and\ \bibinfo {author} {\bibfnamefont {L.}~\bibnamefont
  {You}},\ }\bibfield  {title} {\bibinfo {title} {{Quantum Phase Diffusion of a
  Bose-Einstein Condensate}},\ }\href
  {https://doi.org/10.1103/PhysRevLett.77.3489} {\bibfield  {journal} {\bibinfo
   {journal} {Phys. Rev. Lett.}\ }\textbf {\bibinfo {volume} {77}},\ \bibinfo
  {pages} {3489} (\bibinfo {year} {1996})}\BibitemShut {NoStop}%
\bibitem [{\citenamefont {Lopes}\ \emph {et~al.}(2017)\citenamefont {Lopes},
  \citenamefont {Eigen}, \citenamefont {Navon}, \citenamefont {Cl\'ement},
  \citenamefont {Smith},\ and\ \citenamefont {Hadzibabic}}]{lopes}%
  \BibitemOpen
  \bibfield  {author} {\bibinfo {author} {\bibfnamefont {R.}~\bibnamefont
  {Lopes}}, \bibinfo {author} {\bibfnamefont {C.}~\bibnamefont {Eigen}},
  \bibinfo {author} {\bibfnamefont {N.}~\bibnamefont {Navon}}, \bibinfo
  {author} {\bibfnamefont {D.}~\bibnamefont {Cl\'ement}}, \bibinfo {author}
  {\bibfnamefont {R.~P.}\ \bibnamefont {Smith}},\ and\ \bibinfo {author}
  {\bibfnamefont {Z.}~\bibnamefont {Hadzibabic}},\ }\bibfield  {title}
  {\bibinfo {title} {{Quantum Depletion of a Homogeneous Bose-Einstein
  Condensate}},\ }\href {https://doi.org/10.1103/PhysRevLett.119.190404}
  {\bibfield  {journal} {\bibinfo  {journal} {Phys. Rev. Lett.}\ }\textbf
  {\bibinfo {volume} {119}},\ \bibinfo {pages} {190404} (\bibinfo {year}
  {2017})}\BibitemShut {NoStop}%
\end{thebibliography}%

\widetext
\clearpage
\begin{center}
\textbf{\large Supplementary Material}
\end{center}
\setcounter{equation}{0}
\setcounter{figure}{0}
\setcounter{table}{0}
\setcounter{page}{1}
\makeatletter
\renewcommand{\theequation}{S\arabic{equation}}
\renewcommand{\thefigure}{S\arabic{figure}}
\renewcommand{\bibnumfmt}[1]{[S#1]}
\renewcommand{\citenumfont}[1]{S#1}

\section{Effective theory for an expanding one dimensional BEC}

In order to gain control of all important degrees of freedom in our condensate analogue, in this section we show how a three dimensional gas can behave as an effective system by means of external trapping potentials. Let $\Phi=\Phi(t,\x)$ describe a 3D gas according to the action $(\hbar=1)$
\begin{align}
    S=\int\d^4x\bigg[&\frac{i}{2}(\Phi^*\partial_t\Phi-\Phi\partial_t\Phi^*)-\frac{|\nabla\Phi|^2}{2m}-\left(V_{3D}+\frac{g_{3D}}{2}|\Phi|^2\right)|\Phi|^2\bigg].\label{action1}
\end{align}
Here $V_{3D}$ and $g_{3D}$ are the external potential and the particle interaction strength, respectively.
\begin{figure}[h!]
    \centering
    \includegraphics[width=0.5\linewidth]{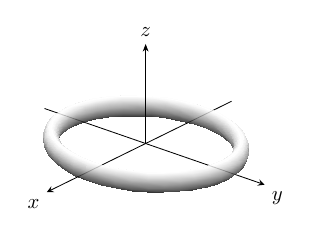}
    \caption{Schematic view of the condensate ring.}
    \label{figsup1}
\end{figure}
By adopting cylindrical coordinates, $\{r,\theta,z\}$, with the convention $-\pi<\theta\leq\pi$, we let $\ell_{r}>0$ be the characteristic size of the gas in the radial and $z$ directions, that is, the gas is confined to the region $|r-R|,|z|<\ell_r$ [see Fig.~\ref{figsup1}], which in turn identifies the torus radius $R$. Also, we assume that $\ell_r\ll R$. Such configuration can be achieved by adjusting the external potential $V_{3D}$ accordingly. As a concrete recipe, let us assume that
\begin{align}
    V_{3D}(t,r,&\theta,z)=V(t,\theta)+\frac{1}{2\ell_r^2m}\left[\frac{\ell_r^2}{4r^2}+\frac{2R}{r}+\frac{(r-R)^2}{\ell_r^2}+\frac{z^2}{\ell_r^2}-4\right],
\end{align}
and we allow $R$ to depend on time. In the limit $\ell_r\rightarrow 0$, we can assume that the field $\Phi$ behaves as $\Phi(t,\x)=\phi(t,\theta)\phi_r(t,r,z)$, where
\begin{equation}
    \phi_r(t,r,z)=\sqrt{\frac{r}{\alpha}}e^{-(r-R)^2/(2\ell_r^2)-z^2/(2\ell_r^2)},
\end{equation}
and $\alpha=\pi\ell_r^2R^2$ is a (possibly time-dependent) normalization constant fixed by
\begin{equation}
    \int_0^\infty\d r r\int_{-\infty}^\infty\d z|\phi_r|^2=1.
\end{equation}
In this regime, the action \eqref{action1} becomes
\begin{align}
    S=\int\d t\d \theta \bigg[&i\phi^*\partial_t\phi-\frac{|\partial_\theta\phi|^2}{2mR^2}-\left(V+\frac{g}{2R}|\phi|^2\right)|\phi|^2\bigg],\label{action2}
\end{align}
where we used that $\ell_r\ll R$, and  $g=g_{3D}/(2\pi\ell_r^2)$ 
%
%
is the effective $1D$ coupling constant. Note that $\ell=2\pi R$ is the condensate size and that the time-dependence of $g$ is inherited from the time-dependence of $g_{3D}$. We adopt Eq.~\eqref{action2} as the effective one dimensional theory for the condensate. By construction, the field $\phi$ is dimensionless.

A key aspect of our analysis is that the above action is invariant under $\phi\rightarrow \exp(i\lambda)\phi$ for real $\lambda$, which implies, by means of the Noether Theorem, that 
\begin{equation}
    \partial_t\rho+\frac{1}{R}\partial_\theta J=0,
\end{equation}
where
\begin{align}
    \rho=|\phi|^2,\label{condensatedensity}\ \ J=\frac{1}{mR}\mbox{Im}(\phi^*\partial_\theta\phi).
\end{align}
In general, the number of particles in the gas,
\begin{equation}
    N=\int_{-\pi}^{\pi}\d\theta\rho,\label{N}
\end{equation}
is time-independent.

The Euler-Lagrange equation associated to \eqref{action2} is the Gross-Pitaevskii (GP) equation
\begin{equation}
    i\partial_t\phi=-\frac{\partial^2_\theta\phi}{2mR^2}+\left(V+\frac{g}{R}|\phi|^2\right)\phi.\label{fieldeq}
\end{equation}
We note also that, by construction, the field $\phi(t,\theta)$ is naturally subjected to the periodic boundary conditions
\begin{align}
\left.\phi\right|_{\theta=-\pi}&=\left.\phi\right|_{\theta=\pi},\\
\left.\partial_\theta\phi\right|_{\theta=-\pi}&=\left.\partial_\theta\phi\right|_{\theta=\pi}.
\end{align}   

We are interested in a scenario for which $g\equiv0$ for $t<0$, and the ring size $\ell\equiv \ell_0$ is time independent, whereas, for $t>0$, $g$ assumes a positive profile and $\d \ell/\d t\equiv\dot{\ell}\neq0$. Let $g_0>0$ be a constant reference value for $g$, and $\mu_0=g_0N/\ell_0$ be the initial chemical potential. In terms of these quantities, Eq.~\eqref{fieldeq} assumes the form
\begin{align}
    \frac{i}{\mu_0}\partial_t\phi=-\frac{\xi_0^2\partial^2_\theta\phi}{2R^2}+\left(\frac{V}{\mu_0}+\frac{g}{g_0}\frac{\ell_0}{R}\frac{|\phi|^2}{N}\right)\phi,
\end{align}
where $\xi_0=1/\sqrt{m\mu_0}$ is the condensate healing length. Henceforth, we adopt units such that distances are expressed in units of $\xi_0$, $t$ is expressed in units of $1/\mu_0$, $V$ is given in units of $\mu_0$, $g$ in units of $g_0$, and $\phi$ in units of $\sqrt{N}$. With these conventions the GP equation assumes the form
\begin{equation}
    i\partial_t\phi=-\frac{\partial^2_\theta\phi}{2R^2}+\left(V+\frac{\ell_0}{R}g|\phi|^2\right)\phi,\label{fieldeq2}
\end{equation}
whereas Eq.~\eqref{N} reads
\begin{equation}
    1=\int_{-\pi}^{\pi}\d \theta\rho.
\end{equation}
%


%
%

\section{Field quantization}

In general, we look for an expansion for the field $\chi$ in terms of a suitable set of field modes. In order to achieve this, let us work with the variable $\psi(t,\theta)=\exp(i\mu t)\chi(t,\theta)$, and the Nambu spinor $\Psi=(\psi,\psi^*)^{T}$, where $T$ is the matrix transpose. In terms of $\Psi$, the BdG equation reads
\begin{align}
    i\sigma_3\partial_t\Psi=-\frac{\partial_\theta^2}{2R^2}\Psi+\frac{R_0 g}{R}(1+\sigma_1)\Psi+\sum_{\mathfrak{m}}\beta_{\mathfrak{m}}e^{i\mathfrak{m} \theta}\left[\frac{R_0g}{R}(1+\sigma_1)-\frac{\mathfrak{m}^2}{4R^2}\right]\Psi+\mathcal{O}(\beta^2),\label{expequation}
\end{align}
where $\sigma_i$, $i=1,2,3$, are the Pauli matrices, and we used the fact that $\beta_{\mathfrak{m}}$ is small. Notice that the Nambu spinor has the symmetry $\Psi=\sigma_1\Psi^*$.

We now proceed to construct a quantum field expansion for $\Psi$. A direct manipulation of the field equation \eqref{expequation} implies that
\begin{equation}
    \langle\Psi,\Psi'\rangle=\int_{-\pi}^\pi\d\theta\Psi^\dagger\sigma_3\Psi',\label{product}
\end{equation}
is time-independent, for
any two solutions $\Psi,\Psi'$ satisfying periodic boundary conditions. We look for a set of positive norm [according to \eqref{product}] solutions $\Psi_\alpha$, where $\alpha$ is some suitable index, normalized as $\langle\Psi_\alpha,\Psi_{\alpha'}\rangle=\delta_{\alpha\alpha'}$. In this case, $\sigma_1\Psi_\alpha^*$ is also a solution, with negative norm: $\langle\sigma_1\Psi_\alpha^*,\sigma_1\Psi_{\alpha'}^*\rangle=-\delta_{\alpha\alpha'}$, and we may expand any spinor $\Psi$ as
\begin{equation}
    \Psi=\sum_{\alpha}(a_\alpha\Psi_\alpha+a_\alpha^*\sigma_1\Psi_\alpha^*).\label{expansion}
\end{equation}
In view of $\langle\sigma_1\Psi_\alpha^*,\Psi_{\alpha'}\rangle=0$, the Fourier coefficients are given by $a_\alpha=\langle\Psi_\alpha,\Psi\rangle$. 

Note that in the absence of anisotropies, the differential operator in Eq.~\eqref{expequation} is independent of $\theta$, and thus the modes can be taken as $\Psi(t,\theta)\equiv\Pi_n^{(0)}(t,\theta)=\Psi^{(0)}_n(t)\exp(in\theta)$, where $n\in\mathbb{Z}$. Accordingly, we find that
\begin{equation}
    \left[i\sigma_3\partial_t-\omega_n-\frac{\gamma}{2}(1+\sigma_1)\right]\Psi^{(0)}_n=0.\label{reducedfieldeq}
\end{equation}
Solutions to this equation in the stationary regime ($t<0$) can be taken as
\begin{equation}
    \Psi_n^{(0)}(t)=\frac{1}{\sqrt{2\pi}}e^{-i\omega_n t}\left(\begin{array}{c}
        1   \\
        0  
    \end{array}\right).\label{psi0initial}
\end{equation}
The particular form of $\Psi_n^{(0)}(t)$ in the expanding period $(t>0)$ therefore depends on $R(t)$ and $g(t)$, and can be calculated by means of the Cauchy problem given by Eq.~\eqref{reducedfieldeq} with the initial condition %
\begin{equation}
    \Psi_n^{(0)}(0)=\frac{1}{\sqrt{2\pi}}\left(\begin{array}{c}
        1   \\
        0  
    \end{array}\right).\label{initialconditionforpsi0}
\end{equation}
Here, $\Psi^{(0)}_n(t)$ for $t<0$ is given by Eq.~\eqref{psi0initial}, whereas for $t>0$ it is the unique solution of the Cauchy problem \eqref{reducedfieldeq}, \eqref{initialconditionforpsi0}.

We note that because we are working perturbatively in the strength of the anisotropies, the solutions for $\Psi$ can be taken in the form
\begin{equation}
    \Psi_n=e^{in\theta}\Psi^{(0)}_n+\sum_\mathfrak{m}\beta_{\mathfrak{m}}e^{i(n+\mathfrak{m})\theta}\Psi^{(1)}_{n,m}+\ldots,
\end{equation}
which, when substituted in Eq.~\eqref{expequation}, leads to
\begin{align}
    \left[i\sigma_3\partial_t-\omega_{n+\mathfrak{m}}-\frac{\gamma}{2}(1+\sigma_1)\right]&\Psi^{(1)}_{n,\mathfrak{m}}=\frac{1}{2}\left[\gamma(1+\sigma_1)-\omega_{\mathfrak{m}}\right]\Psi^{(0)}_n.\label{eqfieldpsi1}
\end{align}
%
%


%
%
%
%
 For $t<0$ (so $\gamma = 0$), the spinors $\Psi^{(1)}_{n,\mathfrak{m}}$ are given by
\begin{align}
   \Psi^{(1)}_{n,\mathfrak{m}}(t)=-\frac{\omega_\mathfrak{m}}{2}\frac{1}{\sqrt{2\pi}}\frac{e^{-i\omega_{n+\epsilon}t}-e^{-i\omega_{n+\mathfrak{m}}t}}{\omega_{n+\epsilon}-\omega_{n+\mathfrak{m}}}\left(\begin{array}{c}
        1   \\
        0  
    \end{array}\right).\label{sigmanonexpanding}
\end{align}
Here $\epsilon>0$ is a small regulator that should be taken to zero at the end of the calculations. It is kept here because if $\mathfrak{m}$ is an even integer number, then $\omega_{n}-\omega_{n+\mathfrak{m}}=0$ when $n=\mathfrak{m}/2$. The physical mechanism behind this is linked to the modification of the eigen frequencies of the $\mathfrak{m}/2$ modes, which is absent in case $\mathfrak{m}$ is odd.  In the expanding period, $\Psi^{(1)}_{n,\mathfrak{m}}(t)$ is the unique solution of \eqref{eqfieldpsi1} with the initial condition fixed by Eq.~\eqref{sigmanonexpanding}.

We find by direct substitution that
\begin{equation}
    \langle\Psi_n,\Psi_{n'}\rangle=\delta_{nn'}+\mathcal{O}(\beta^2),
\end{equation}
and the set $\{\Psi_n\}_{n\in\mathbb{Z}}$ is a complete set of positive norm field modes.

Quantization is then finished by promoting each of the Fourier coefficients $a_n$ in Eq.~\eqref{expansion} to an operator subjected to the commutation relation $[a_n,a^\dagger_{n'}]=\delta_{nn'}$, and we identify the state $|0\rangle$ defined by $a_n|0\rangle=0$, for all $n$, with the system vacuum state. If $\Psi$ is an arbitrary spinor, we denote its components by $\Psi_{i}$, $i=1,2$, and thus we obtain the quantum field expansion for $\chi$ by taking the first component of Eq.~\eqref{expansion}:
\begin{equation}
    \chi(t,\theta)=e^{-i\mu t}\sum_{n\in\mathbb{Z}}\left[a_n\Psi_{n,1}(t,\theta)+a^\dagger_n\Psi^*_{n,2}(t,\theta)\right],\label{quantumfieldexpansion}
\end{equation}
which, by construction, satisfies the canonical commutation relations.

\subsection{A case where exact solutions can always be found}

It should be mentioned that finding analytical solutions for Eqs.~\eqref{reducedfieldeq} and 
\eqref{eqfieldpsi1} might not be possible due to the fact that both $\omega_n$ and $\gamma$ depend on time in an arbitrary manner, through $R(t),g(t)$. For the sake of completeness and in order to produce a reference solution to be compared with numerical data, we now show that there is a family of analogues for which the field modes assume a simple form for arbitrary $R(t)$. In fact, recall that, at $t=0$, the particle interactions are turned on, i.e., $g=1$, and the condensate expansion takes place. Let us suppose that, for $t>0$, $gR=R_0$. In this particular regime, $R^2\omega_n$ and $R^2\gamma$ remain time-independent, and, by working with the variable $\tau$ defined as
\begin{equation}
    \tau=\int_0^{t}\frac{\d t'}{R^2(t')},
\end{equation}
equation \eqref{reducedfieldeq}, when expressed in terms of $\tau$ derivatives, becomes stationary, and as such can be solved in a straightforward manner. For instance, if $n\neq0$,
\begin{equation}
    e^{-i\Omega_n\tau}\Gamma_n,
\end{equation}
with the time-independent quantities, 
\begin{align}
    \Omega_n&=R^2\sqrt{\omega_n(\omega_n+\gamma)},\\
    \Gamma_n&=\left(\begin{array}{c}
         R^2\gamma/2  \\
         \Omega_n-R^2\omega_n-R^2\gamma/2 
    \end{array}\right),
\end{align}
solves Eq.~\eqref{reducedfieldeq} for $t>0$, and a second linearly independent solution is $\exp(i\Omega_n\tau)\tilde{\Gamma}_n$, with $\tilde{\Gamma}_n=\sigma_1\Gamma_n^*$. When $n=0$, two solutions for Eq.~\eqref{reducedfieldeq} are
\begin{align}
    \Gamma_0&=\frac{1}{2}\left(\begin{array}{c}
         1  \\
         -1 
    \end{array}\right),\\
    \tilde{\Gamma}_0&=\frac{1}{2}\left(\begin{array}{c}
         1  \\
         1 
    \end{array}\right)-i\frac{R^2\gamma \tau}{2}\left(\begin{array}{c}
         1  \\
         -1 
    \end{array}\right).
\end{align}
Therefore, $\Psi^{(0)}_n(t)$ during the expansion period must take the form
\begin{equation}
    \Psi^{(0)}_n(t)=\eta_{1,n}e^{-i\Omega_n\tau}\Gamma_n+\eta_{2,n}e^{i\Omega_n\tau}\tilde{\Gamma}_n,\label{psizero}
\end{equation}
and the coefficients $\eta_{i,n}$ are found from the initial conditions to be
\begin{align}
    \eta_{1,n}&=\frac{1}{\sqrt{2\pi}}\frac{\tilde{\Gamma}^{T}_n\sigma_2(1,0)^{T}}{\tilde{\Gamma}^{T}_n\sigma_2\Gamma_n},\\
    \eta_{2,n}&=\frac{1}{\sqrt{2\pi}}\frac{\Gamma^{T}_n\sigma_2(1,0)^{T}}{\Gamma^{T}_n\sigma_2\tilde{\Gamma}_n}.
\end{align}

In a similar manner, we can write down the solution of Eq.~\eqref{eqfieldpsi1} as follows. By defining the time-dependent matrix
\begin{align}
    \mathcal{M}_{n,\mathfrak{m}}=&\frac{1}{{\Omega^2_{n+\epsilon}-\Omega^2_{n+\mathfrak{m}}}}\bigg\{e^{-i\Omega_{n+\epsilon}\tau}\left[\Omega_{n+\epsilon}\sigma_3+R^2\omega_{n+\mathfrak{m}}+\frac{R^2\gamma}{2}(1-\sigma_1)\right]\nonumber\\
    &-e^{-i\Omega_{n+\mathfrak{m}}\tau}\left[\Omega_{n+\mathfrak{m}}\sigma_3+R^2\omega_{n+\mathfrak{m}}+\frac{R^2\gamma}{2}(1-\sigma_1)\right]\bigg\}
    , 
\end{align}
we note that the identity
\begin{align}
    e^{-i\Omega_{n+\epsilon}\tau}&=\left[i\sigma_3\partial_\tau-R^2\omega_{n+\mathfrak{m}}-\frac{R^2\gamma}{2}(1+\sigma_1)\right]\mathcal{M}_{n,\mathfrak{m}}
\end{align}
holds for all $\epsilon$. With that it follows that the most general solution of Eq.~\eqref{eqfieldpsi1} with $\Psi^{(0)}_n$ given by Eq.~\eqref{psizero} is
\begin{align}
    \Psi^{(1)}_{n,\mathfrak{m}}(t)=&\nu_{1,n,\mathfrak{m}}e^{-i\Omega_{n+\mathfrak{m}}\tau}\Gamma_{n+\mathfrak{m}}+\nu_{2,n,\mathfrak{m}}e^{i\Omega_{n+\mathfrak{m}}\tau}\tilde{\Gamma}_{n+\mathfrak{m}}\nonumber\\
    &+\frac{\eta_{1,n}}{2}\mathcal{M}_{n,\mathfrak{m}}[\gamma(1+\sigma_1)-\omega_{\mathfrak{m}}]\Gamma_n+\frac{\eta_{2,n}}{2}\sigma_1\mathcal{M}^{*}_{n,\mathfrak{m}}\sigma_1[\gamma(1+\sigma_1)-\omega_{\mathfrak{m}}]\tilde{\Gamma}_n.\label{solforsigma}
\end{align}
%
%
%
The terms with $\eta_{i,n}$ in Eq.~\eqref{solforsigma} give rise to a particular solution for Eq.~\eqref{eqfieldpsi1}, whereas the terms with $\nu_{i,n,\mathfrak{m}}$ are solutions to the homogeneous part of the equation. Finally, the coefficients $\nu_{i,n,\mathfrak{m}}$ are fixed by the initial conditions on $\Psi^{(1)}_{n,\mathfrak{m}}$ at $t=0$ [cf.~\eqref{sigmanonexpanding}]. We find that
\begin{align}
    \nu_{1,n,\mathfrak{m}}=&-\frac{\eta_{1,n}}{2}\frac{\tilde{\Gamma}^T_{n+\mathfrak{m}}\sigma_2\mathcal{M}_{n,\mathfrak{m}}[\gamma(1+\sigma_1)-\omega_{\mathfrak{m}}]\Gamma_n}{\tilde{\Gamma}^T_{n+\mathfrak{m}}\sigma_2\Gamma_{n+\mathfrak{m}}}-\frac{\eta_{2,n}}{2}\frac{\tilde{\Gamma}^T_{n+\mathfrak{m}}\sigma_2\sigma_1\mathcal{M}^{*}_{n,\mathfrak{m}}\sigma_1[\gamma(1+\sigma_1)-\omega_{\mathfrak{m}}]\tilde{\Gamma}_n}{\tilde{\Gamma}^T_{n+\mathfrak{m}}\sigma_2\Gamma_{n+\mathfrak{m}}},\nonumber\\
    \nu_{2,n,\mathfrak{m}}=&-\frac{\eta_{1,n}}{2}\frac{\Gamma^T_{n+\mathfrak{m}}\sigma_2\mathcal{M}_{n,\mathfrak{m}}[\gamma(1+\sigma_1)-\omega_{\mathfrak{m}}]\Gamma_n}{\Gamma^T_{n+\mathfrak{m}}\sigma_2\tilde{\Gamma}_{n+\mathfrak{m}}}-\frac{\eta_{2,n}}{2}\frac{\Gamma^T_{n+\mathfrak{m}}\sigma_2\sigma_1\mathcal{M}^{*}_{n,\mathfrak{m}}\sigma_1[\gamma(1+\sigma_1)-\omega_{\mathfrak{m}}]\tilde{\Gamma}_n}{\Gamma^T_{n+\mathfrak{m}}\sigma_2\tilde{\Gamma}_{n+\mathfrak{m}}},
\end{align}
where all the functions in the above are evaluated at $t=0$. This finishes the construction of an exact solution for the quantum field expansion.

\end{document}